\documentclass[11pt]{article}
\usepackage{amssymb,amsmath,amsfonts}
\usepackage{graphicx}
\usepackage{graphics}
\usepackage{eepic,epsfig}
\usepackage{dcolumn}
\usepackage{color}

\@addtoreset{equation}{section}

\makeatother

\begin{document}

\title{Vacuum fermionic currents in braneworld models\\
on AdS bulk with a cosmic string}
\author{S. Bellucci$^1$, W. Oliveira dos Santos$^{1,2}$, E. R. Bezerra de
Mello$^2$, A. A. Saharian$^3$ \vspace{0.3cm} \\
\\
\textit{$^1$INFN, Laboratori Nazionali di Frascati,}\\
\textit{Via Enrico Fermi 40, 00044 Frascati, Italy} \vspace{0.3cm}\\
\textit{$^2$Departamento de F\'{\i}sica, Universidade Federal da Para\'{\i}%
ba,} \\
\textit{58059-900, Caixa Postal 5008, Jo\~{a}o Pessoa, PB, Brazil} \vspace{%
0.3cm}\\
\textit{$^3$ Department of Physics, Yerevan State University,}\\
\textit{1 Alex Manoogian Street, 0025 Yerevan, Armenia }}
\maketitle

\begin{abstract}
We investigate the effects of a brane and magnetic-flux-carrying cosmic
string on the vacuum expectation value (VEV) of the current density for a
charged fermionic field in the background geometry of (4+1)-dimensional
anti--de Sitter (AdS) spacetime. The brane is parallel to the AdS boundary
and the cosmic string is orthogonal to the brane. Two types of boundary
conditions are considered on the brane that include the MIT bag boundary
condition and the boundary conditions in $Z_{2}$-symmetric braneworld
models. The brane divides the space into two regions with different
properties of the vacuum state. The only nonzero component of the current
density is along the azimuthal direction and in both the regions the
corresponding VEV is decomposed into the brane-free and brane-induced
contributions. The latter vanishes on the string and near the string the
total current is dominated by the brane-free part. At large distances from
the string and in the region between the brane and AdS horizon the decay of
the brane-induced current density, as a function of the proper distance, is
power-law for both massless and massive fields. For a massive field this
behavior is essentially different from that in the Minkowski bulk. In the
region between the brane and AdS boundary the large-distance decay of the
current density is exponential. Depending on the boundary condition on the
brane, the brane-induced contribution is dominant or subdominant in the
total current density at large distances from the string. By using the
results for fields realizing two inequivalent irreducible representations of
the Clifford algebra, the vacuum current density is investigated in $C$- and
$P$-symmetric fermionic models. Applications are given for a cosmic string
in the Randall-Sundrum-type braneworld model with a single brane.
\end{abstract}

\bigskip

Keywords: Cosmic string, AdS space, Brane, Vacuum polarization

\bigskip

\section{Introduction}

\label{Int}

The anti-de Sitter (AdS) spacetime is one of the most interesting solutions
of General Relativity. The relevance of this maximally symmetric geometry
from the point of view of quantum field theory resides in the fact that a
large number of field theoretical problems are exactly soluble on its
background. This allows to reveal information on the influence of
gravitational field on quantum matter in less symmetric geometries. In
addition, the length scale related to the AdS constant curvature, can serve
as a regularization parameter for infrared divergences in interacting
quantum field theories without reducing the number of symmetries \cite%
{Callan1990}. The AdS spacetime has been the subject of considerable
attention due to the realization that it emerges as a stable ground state
solution in Kaluza-Klein models, extended supergravity and string theories
and approximates the near-horizon geometries of extremal black holes, domain
walls and $p$-brane configurations in string theories. In addition, the AdS
background geometry plays a crucial role in the AdS/CFT correspondence and
braneworlds models with large extra dimensions. These two exciting
developments of modern theoretical physics naturally appear in the
string/M-theory context and offer an interesting framework to approach
several problems in particle physics, cosmology and condensed matter
physics. Motivated by the points given above, quantum field theory in AdS
background has been extensively studied by several authors (see, for
instance, \cite{Fronsdal1974}-\cite{Morl20} and references therein).

In the present paper we investigate the effects of a cosmic string and brane
on the vacuum expectation value (VEV) of the fermionic current density
induced by a magnetic flux tube in AdS spacetime. Cosmic strings are among
the most interesting types of topological defects that might have formed as
a consequence of spontaneous symmetry breaking at phase transitions in the
early Universe \cite{Kibble,V-S,Hind95}. As possible seeds for large scale
structure formation in the Universe, they have been extensively studied in
the eighties and early nineties of the last century. Although the
observational data on the temperature anisotropies of the cosmic microwave
background radiation (CMB) supported the inflationary scenario for the
density perturbations, excluding the cosmic strings as the main origin of
structures, they are still sources for a number of interesting physical
effects. The latter include the gravitational lensing, creation of small
non-Gaussianities in the CMB, the generation of gravitational waves, gamma
ray bursts and high-energy cosmic rays.

Initially, cosmic strings have been regarded as completely different from
fundamental superstrings. In string theory the latter are considered as the
basic building blocks of matter. A particularly interesting topic in recent
developments of the subject is the investigation of possibility for
appearance of low tension cosmic strings in string theory (see \cite{Witt85}
for early discussion). Models were proposed where, under certain conditions,
the fundamental strings can grow to macroscopic size with features similar
to those for cosmic strings. Among the examples of this kind of models is
the brane-inflation scenario \cite{Dvali1999} (for reviews see \cite%
{Tye2008,Cope10,Cope11,Cher15}). In this scenario the inflation appears
naturally in the braneworld context and it is a specific realization of the
inflationary paradigm in the early Universe within the braneworld picture in
the string theory. Among other interesting predictions in this theoretical
framework, the production of cosmic strings towards the end of inflation has
attracted considerable attention \cite{Sarangi2002}. It has been suggested
that the detection of cosmic strings could be a possible way to verify the
inflationary paradigm, providing an interesting testing ground into string
theory by probing the braneworld scenario before the inflationary phase \cite%
{Tye2008,Cope10,Cope11,Cher15}. Many other publications have investigated
the effects of braneworld gravity on the physics of cosmic strings \cite%
{Davis2001,Davis2007,Heydari-Fard,Braga2005,Abdalla2007,Abdalla2015}.
Specifically in \cite{Abdalla2015}, the authors have studied the
consequences of the braneworld modified gravity on the cosmic string
phenomenology. By considering the presence of a cosmic string they were able
to put a strong constraint on the braneworld tension, making it compatible
with the observational data for temperature anisotropies of CMB. They also
have shown that the deficit angle induced in the string's outside region is
significantly attenuated due to the braneworld gravitational corrections,
providing a possible explanation for the null experimental results based on
the cosmic string effects caused by its conical geometry, such as lensing
effects. This close relation between braneworlds and cosmic strings gives us
enough motivation to consider both objects within a unified picture for
studying vacuum quantum effects in the present work.

The geometry associated with a cosmic string in AdS spacetime has been
analyzed in \cite{Ghe1,Cristine}. There the authors have shown that, similar
to the case of Minkowskian spacetime, at distances from the string larger
than its core radius, the gravitational effects of the string can be
approximated by a planar angle deficit in the two-dimensional subspace
orthogonal to the string in the AdS line-element. The combination of
non-trivial topology and the spacetime curvature induce vacuum polarization
effects for quantum fields. In this sense the investigation of the influence
of the presence of a cosmic string in AdS bulk on the vacuum polarization of
quantum fields, have been investigated for scalar and fermionic fields in
\cite{deMell:2011ji} and \cite{Beze13}, respectively. Taking into account
the presence of a magnetic flux along the string core, in \cite{Oliv19} the
vacuum current for a charged scalar field in AdS bulk was investigated. The
VEV of the corresponding energy-momentum tensor was analyzed in \cite{Oliv20}%
. Finally, in \cite{Oliveira} the induced fermionic current in the AdS bulk
in the presence of a cosmic string was discussed.\footnote{%
In fact, in references \cite{Oliv19}, \cite{Oliv20} and \cite{Oliveira},
higher dimensional AdS spacetimes were considered, admitting that an extra
dimension coordinate was compactified.}

Here we consider the VEV of the fermionic current density around
magnetic-flux-carrying cosmic string on AdS bulk in the presence of a brane
parallel to the AdS boundary. The cosmic string is orthogonal to the brane
and on the brane the fermionic field is constrained by two types of boundary
conditions. The first one corresponds to the MIT bag boundary condition and
the second one is obtained from the bag boundary condition changing the sign
of the term with the normal to the boundary. In braneworld models these
boundary conditions are dictated by the $Z_{2}$-symmetry under the
reflection with respect to the brane. Note that the vacuum current densities
on locally AdS bulk with a part of spatial dimensions compactified to a
torus and in the presence of branes have been investigated in \cite%
{Beze15a,Bell15,Bell16} and \cite{Bell17,Bell18,Bell20} for charged scalar
and fermionic fields, respectively.

The paper is organized as follows. In section \ref{sec:Setup} we describe
the setup of the problem and present the complete set of positive and
negative energy solutions to the Dirac equation in the presence of a brane
parallel to the AdS boundary. In section \ref{sec:Rreg}, the VEV of the
current density in the region between the brane and the AdS horizon
(R-region) is investigated. Various asymptotic limits are considered and
numerical results are presented. Similar investigations for the region
between the brane and AdS boundary (L-region) are presented in section \ref%
{sec:Lreg}. In section \ref{sec:BC2} we consider the second type of boundary
condition and analyse the effects of the brane on the current densities in
the R- and L-regions. Discussion of our results in the context of $C$- and $%
P $-symmetric models is presented in section \ref{sec:CP}. In section \ref%
{sec:RS}, an application to the Randall-Sundrum model with a single brane is
provided. Section \ref{sec:Conc} summarizes the most relevant results
obtained. Throughout the paper, we use natural units $G=\hbar =c=1$.

\section{Problem setup and the fermionic modes}

\label{sec:Setup}

We consider a fermionic quantum field $\psi (x)$ in background of $(4+1)$%
-dimensional spacetime with the metric tensor $g_{\mu \nu }$ defined by the
line element
\begin{equation}
ds^{2}=g_{\mu \nu }dx^{\mu }dx^{\nu }=\left( \frac{a}{w}\right) ^{2}\left(
dt^{2}-dr^{2}-r^{2}d\phi ^{2}-dw^{2}-dz^{2}\right) \ ,  \label{ds2}
\end{equation}%
where $-\infty <t,z<+\infty $, $r\geq 0$, $0\leq \phi \leq \phi _{0}$, $%
0\leq w<\infty $. For $\phi _{0}=2\pi $, Eq. (\ref{ds2}) presents the AdS
spacetime described in Poincar\'{e} coordinates (with polar coordinates $%
(r,\phi )$ in two-dimensional subspace). The parameter $a$ defines the
cooresponding curvature radius and is related to negative cosmological
constant $\Lambda $ and the Ricci scalar $R$ by formulas $\Lambda =-6/a^{2}$
and $R=-20/a^{2}$. The limiting values of the $w$-coordinate, $w=0$ and $%
w=\infty $, determine the AdS boundary and the AdS horizon, respectively. In
what follows we will be concerned with the case $\phi _{0}<2\pi $ that
corresponds to two-dimensional conical subspace $(r,\phi )$ with planar
angle deficit $2\pi -\phi _{0}$. Note that the latter does not change the
local geometry for $r\neq 0$. However, the topology is changed and this
gives rise interesting effects in quantum field theory. The line element (%
\ref{ds2}) without the term $dz^{2}$ describes an idealized cosmic string in
background of AdS spacetime.

In the presence of a classical gauge field $A_{\mu }$, the Dirac equation
for the field $\psi (x)$ reads
\begin{equation}
\left[ i\gamma ^{\mu }(\partial _{\mu }+\Gamma _{\mu }+ieA_{\mu })-sm\right]
\psi =0,  \label{Direq}
\end{equation}%
where $\Gamma _{\mu }$ is the spin connection. In $(4+1)$-dimensional
spacetime the irreducible representation of the Clifford algebra is realized
by $4\times 4$ Dirac matrices $\gamma ^{\mu }$. In odd spacetime dimensions
there are two inequivalent irreducible representations and the parameter $%
s=\pm 1$ corresponds to those representations (see the discussion in section %
\ref{sec:CP} below). We assume the presence of a codimension one brane
parallel to the AdS boundary and located at $w=w_{0}$. On the brane the
field operator is constrained by the MIT bag boundary condition
\begin{equation}
(1+i\gamma ^{\mu }n_{\mu })\psi (x)=0\ ,\ \ w=w_{0}\ ,  \label{MITbc}
\end{equation}%
where $n_{\mu }$ is the corresponding normal vector. In terms of the
coordinate $w$ the line element (\ref{ds2}) is written in conformally flat
form for $r\neq 0$. It is conformally related to the line element associated
with a cosmic string in $(4+1)$-dimensional Minkowski spacetime. The
physical distance from the brane is measured in terms of the coordinate $y$,
defined as $y=a\ln (w/a)$, $-\infty <y<+\infty $. For the $y$-coordinate of
the brane one has $y_{0}=a\ln (w_{0}/a)$. The distance from the brane is
expressed as $|y-y_{0}|$. The brane separates the background space into two
regions: $0\leq w\leq w_{0}$ and $w\geq w_{0}$. We shall refer them as
L(left)- and R(right)- regions, respectively. In the coordinate system $%
x^{\mu }=(t,r,\phi ,w,z)$, for the normal in the boundary condition (\ref%
{MITbc}) one has $n_{\mu }=\delta _{\mu }^{3}a/w$ in the L-region and $%
n_{\mu }=-\delta _{\mu }^{3}a/w$ in the R-region.

In the discussion below, the gamma matrices will be taken in the following
representation:
\begin{equation}
\gamma ^{0}=\frac{w}{a}\left( {%
\begin{array}{cc}
0 & -i \\
i & 0%
\end{array}%
}\right) \ ,\quad \gamma ^{l}=-i\frac{w}{a}\left( {%
\begin{array}{cc}
\sigma ^{l} & 0 \\
0 & -\sigma ^{l}%
\end{array}%
}\right) \ ,\quad \gamma ^{4}=\frac{w}{a}\left( {%
\begin{array}{cc}
0 & i \\
i & 0%
\end{array}%
}\right)  \label{gam}
\end{equation}%
where $l=1,2,3$ correspond to the coordinates $r,\phi ,w$. In the respective
expressions $2\times 2$ matrices $\sigma ^{l}$ are defined as
\begin{equation}
\sigma ^{1}=\left( {%
\begin{array}{cc}
0 & e^{-iq\phi } \\
e^{iq\phi } & 0%
\end{array}%
}\right) \ ,\ \sigma ^{2}=\frac{i}{r}\left( {%
\begin{array}{cc}
0 & -e^{-iq\phi } \\
e^{iq\phi } & 0%
\end{array}%
}\right) \ ,  \label{Pauli}
\end{equation}%
with $q=2\pi /\phi _{0}$, and $\sigma ^{3}=\mathrm{diag}(1,-1)$. These are
the Pauli matrices in the coordinate system $(r,\phi ,w)$. With this choice
of the gamma matrices, the product $\gamma ^{\mu }\Gamma _{\mu }$ in the
field equation (\ref{Direq}) takes the form
\begin{equation}
\gamma ^{\mu }\Gamma _{\mu }=\frac{1-q}{2r}\gamma ^{1}-\frac{2}{w}\gamma
^{3}.  \label{gamGam}
\end{equation}%
In the absence of the cosmic string one has $q=1$.

Our interest in the present paper is the VEV of the fermionic current
density $j^{\mu }=e\bar{\psi}\gamma ^{\mu }\psi $, with $\bar{\psi}=\psi
^{\dagger }\gamma ^{(0)}$ being the Dirac adjoint defined in terms of the
flat spacetime gamma matrix $\gamma ^{(0)}=(a/w)\gamma ^{0}$. This VEV $%
\left\langle 0\right\vert j^{\mu }\left\vert 0\right\rangle \equiv
\left\langle j^{\mu }\right\rangle $ can be presented as a mode sum over
complete set of positive and negative energy fermionic wave-functions $%
\{\psi _{\sigma }^{(+)},\psi _{\sigma }^{(-)}\}$, where the set of quantum
numbers specifies the modes. Expanding the field operators in the expression
for $j^{\mu }$ over the functions $\psi _{\sigma }^{(\pm )}$ and by using
the definition of the vacuum state one can see that
\begin{equation}
\langle j^{\mu }\rangle =\frac{e}{2}\sum_{\sigma }[\bar{\psi}_{\sigma
}^{(-)}\gamma ^{\mu }\psi _{\sigma }^{(-)}-\bar{\psi}_{\sigma }^{(+)}\gamma
^{\mu }\psi _{\sigma }^{(+)}]\ .  \label{jmu}
\end{equation}%
For the gauge field in (\ref{Direq}) we will assume a simple configuration $%
A_{\mu }=\delta _{\mu }^{2}A_{2}$ with constant $A_{2}$. This corresponds to
a magnetic flux along the string's core. Hence, for evaluation of the vacuum
current density one needs complete set of solutions to (\ref{Direq}) for the
special case of the vector potential and obeying the boundary condition (\ref%
{MITbc}).

The general procedure for solving the Dirac equation is similar to that
already discussed in \cite{Beze13} for (3+1)-dimensional AdS spacetime and
in \cite{Oliveira} for the problem on (4+1)-dimensional AdS bulk with
compactified spatial dimension. Here we will describe the main steps only.
From the problem symmetry the dependence of the mode functions on the
spacetime coordinates $t$ and $z$ can be taken in the standard plane wave
form $e^{ik_{z}z\mp iEt}$, where $E>0$ is the energy and the upper and lower
signs correspond to the positive and negative energy modes. Decomposing the
four-component spinor $\psi $ into two-component spinors $\varphi _{\uparrow
}$ and $\varphi _{\downarrow }$, $\psi =(\varphi _{\uparrow },\varphi
_{\downarrow })^{T}$, from the Dirac equation (\ref{Direq}) we obtain two
coupled first order differential equations for the upper and lower
components. From that system of equations the second order differential
equations are derived for the separate functions. Separating the variables $%
r $ and $w$ in both these equations, the corresponding functions are
expressed in terms of the cylinder functions. Then, additional conditions,
previously discussed in \cite{Bordag}, are imposed that relate different
components of the spinor. As a result all those manipulations, the positive
and negative energy fermionic mode functions are presented as%
\begin{equation}
\psi _{\sigma }^{(\pm )}(x)=C_{\sigma }^{(\pm )}w^{5/2}\left( {%
\begin{array}{c}
J_{\beta _{j}}(\lambda r)W_{\nu _{1}}(pw)e^{-iq\phi /2} \\
-\epsilon _{j}\kappa _{\eta }b_{\eta }^{(+)}J_{\beta _{j}+\epsilon
_{j}}(\lambda r)W_{\nu _{2}}(pw)e^{iq\phi /2} \\
i\kappa _{\eta }J_{\beta _{j}}(\lambda r)W_{\nu _{2}}(pw)e^{-iq\phi /2} \\
i\epsilon _{j}b_{\eta }^{(+)}J_{\beta _{j}+\epsilon _{j}}(\lambda r)W_{\nu
_{1}}(pw)e^{iq\phi /2}%
\end{array}%
}\right) e^{iqj\phi +ik_{z}z\mp iEt}\ ,  \label{FermMod}
\end{equation}%
with $0\leq \lambda ,p<\infty $, $j=\pm 1/2,\pm 3/2,\ldots $, and
\begin{equation}
E=\sqrt{\lambda ^{2}+p^{2}+k_{z}^{2}}.  \label{E}
\end{equation}%
The orders of the functions are given by the expressions%
\begin{equation}
\beta _{j}=q|j+\alpha |-\epsilon _{j}/2\ ,\;\nu _{l}=ma+(-1)^{l}s/2,
\label{n12}
\end{equation}%
where $\alpha =eA_{2}/q$, $\epsilon _{j}=1$ for $j>-\alpha $ and $\epsilon
_{j}=-1$ for $j<-\alpha $. In (\ref{FermMod}),
\begin{equation}
W_{\nu }(pw)=c_{1}J_{\nu }(pw)+c_{2}Y_{\nu }(pw),  \label{Wnu}
\end{equation}%
where $J_{\nu }(x)$ and $Y_{\nu }(x)$ are the Bessel and Neumann functions
and we have introduced the notations%
\begin{eqnarray}
\kappa _{\eta } &=&\frac{-k_{z}+\eta \sqrt{k_{z}^{2}+p^{2}}}{p},\ \eta =\pm
1\ ,  \notag \\
b_{\eta }^{(\pm )} &=&\frac{E\mp \eta \sqrt{k_{z}^{2}+p^{2}}}{\lambda }.
\label{bs}
\end{eqnarray}%
Note that one of the constants $c_{1}$ and $c_{2}$ in (\ref{Wnu}) can be
absorbed in the coefficient $C_{\sigma }^{(\pm )}$. With the definitions
given above, the set of quantum numbers $\sigma $ in (\ref{jmu}) is
specified as $\sigma =(\lambda ,p,j,k_{z},\eta )$. It can be checked that
the functions (\ref{FermMod}) are the eigenfunctions of the $w$-projection
of the total angular momentum $J_{3}$ with the eigenvalues $qj$:%
\begin{equation}
\hat{J}_{3}\psi _{\sigma }^{(\pm )}(x)=\left( -i\partial _{\phi }+\frac{q}{2}%
\Sigma ^{3}\right) \psi _{\sigma }^{(\pm )}(x)=qj\psi _{\sigma }^{(\pm )}(x),
\label{J3eig}
\end{equation}%
where $\Sigma ^{3}=\mathrm{diag}(\sigma ^{3},\sigma ^{3})$. The parameter $%
\alpha $ in the definition of $\beta _{j}$ is expressed in terms of the
magnetic flux $\Phi $ along the string axis by the relation $\alpha =-\Phi
_{\phi }/\Phi _{0}$, where $\Phi _{0}=2\pi /e$ is the quantum of the flux.

The coefficient $C_{\sigma }^{(\pm )}$ in (\ref{FermMod}) is determined from
the normalization condition
\begin{equation}
\int_{0}^{\infty }dr\int_{0}^{\phi _{0}}d\phi \int_{-\infty }^{+\infty
}dz\int dw\,r(a/w)^{4}(\psi _{\sigma }^{(\pm )})^{\dagger }\psi _{\sigma
^{\prime }}^{(\pm )}=\delta _{\sigma \sigma ^{\prime }}\ ,\   \label{nc}
\end{equation}%
where $\delta _{\sigma \sigma ^{\prime }}$ is understood as the Dirac delta
function for continuous components of $\sigma $ and the Kronecker delta for
discrete ones. The integration over $w$ in (\ref{nc}) goes over $[0,w_{0}]$
in the L-region and over $[w_{0},\infty )$ in the R-region. The remaining
coefficient in the linear combination is determined from the boundary
condition (\ref{MITbc}).

Note that along the azimuthal direction the mode functions (\ref{FermMod})
obey the periodicity condition $\psi _{\sigma }^{(\pm )}(t,r,\phi +\phi
_{0},w,z)=\psi _{\sigma }^{(\pm )}(t,r,\phi ,w,z)$. We could consider a more
general quasiperiodicity condition $\psi (t,r,\phi +\phi _{0},w,z)=e^{2\pi
i\chi }\psi (t,r,\phi ,w,z)$ with a constant phase $2\pi \chi $. By the
gauge transformation $A_{\mu }^{\prime }=A_{\mu }+\delta _{\mu }^{2}q\chi /e$%
, $\psi ^{\prime }(x)=\psi (x)e^{-iq\chi \phi }$, the set of the fields $%
(A_{\mu },\psi (x))$ is transformed to the set $(A_{\mu }^{\prime },\psi
^{\prime }(x))$, where the new field $\psi ^{\prime }(x)$ is periodic along
the azimuthal direction. From here we conclude that the VEV of the current
density for the field obeying the quasiperiodicity condition with the phase $%
2\pi \chi $ is obtained from the corresponding results given below for
periodic fields by the replacement $\alpha \rightarrow \alpha +\chi $.

The properties of the vacuum state are different in the L- and R-regions and
separate consideration for the fermionic current is required. We start our
investigation with the R-region.

\section{Current density in the R-region}

\label{sec:Rreg}

In this section we consider the current density in the region $w>w_{0}$
(R-region).

\subsection{General formula}

The corresponding modes are given by (\ref{FermMod}), where the function $%
W_{\nu }(pw)$ is defined as (\ref{Wnu}). From the boundary condition (\ref%
{MITbc}) it follows that the coefficients in the corresponding linear
combination can be taken as $c_{1}=Y_{\nu _{2}}(pw_{0})$ and $c_{2}=-J_{\nu
_{2}}(pw_{0})$. Hence, in the R-region the mode functions have the form (\ref%
{FermMod}), where%
\begin{equation}
W_{\nu _{l}}(pw)=G_{\nu _{2},\nu _{l}}(pw_{0},pw),\;l=1,2.  \label{WnuR}
\end{equation}%
Here and in what follows we use the notation
\begin{equation}
G_{\mu ,\nu }(x,y)=Y_{\mu }(x)J_{\nu }(y)-J_{\mu }(x)Y_{\nu }(y).  \label{ge}
\end{equation}%
The eigenvalues of the quantum number $p$ are continuous and in the
normalization condition (\ref{nc}) the corresponding part in the right-hand
side is understood as the delta function $\delta (p-p^{\prime })$. In the
evaluation of the normalization integral we note that for $p=p^{\prime }$
the dominant contribution to the integral over $w$ comes from large values
of $w$ and we can replace the cylinder functions with the arguments $pw$ by
the corresponding asymptotic expressions. In this way one can see that $%
C_{\sigma }^{(\pm )}=C_{\mathrm{(R)}\sigma }^{(\pm )}$ with
\begin{equation}
\left\vert C_{\mathrm{(R)}\sigma }^{(\pm )}\right\vert ^{-2}=\frac{16\pi
^{2}a^{4}\eta }{q\lambda ^{2}p^{2}}E\sqrt{E^{2}-\lambda ^{2}}\kappa _{\eta
}b_{\eta }^{(\pm )}\left[ J_{\nu _{2}}^{2}(pw_{0})+Y_{\nu _{2}}^{2}(pw_{0})%
\right] ,  \label{CR}
\end{equation}%
and with definitions from (\ref{bs}).

Given the mode functions, the VEV of the current density in the R-region is
evaluated by using the formula (\ref{jmu}), where%
\begin{equation}
\sum_{\sigma }=\int_{0}^{\infty }d\lambda \int_{-\infty }^{+\infty
}dk_{z}\int_{0}^{\infty }dp\sum_{\eta =\pm 1}\sum_{j=\pm 1/2,...}.
\label{Rsig}
\end{equation}%
First of all, we can see that for the mode functions (\ref{FermMod}) $\bar{%
\psi}_{\sigma }^{(\pm )}\gamma ^{1}\psi _{\sigma }^{(\pm )}=\bar{\psi}%
_{\sigma }^{(\pm )}\gamma ^{3}\psi _{\sigma }^{(\pm )}=0$ and consequently $%
\langle j^{1}\rangle =\langle j^{3}\rangle =0$. For the components $\mu =0,4$
it can be seen that the product $\bar{\psi}_{\sigma }^{(\pm )}\gamma ^{\mu
}\psi _{\sigma }^{(\pm )}$ is an odd function of $k_{z}$. By taking into
account that in the formula for the VEV\ of the current density one has
integration $\int_{-\infty }^{+\infty }dk_{z}$, we conclude that $\langle
j^{0}\rangle =\langle j^{4}\rangle =0$. Hence, the only nonzero component
corresponds to the current density along the azimuthal direction, $\langle
j^{2}\rangle $. The terms in the mode sum with $\eta =+1$ and $\eta =-1$
give the same contribution and, after summation over $\eta $, the
corresponding physical component $\langle j_{\phi }\rangle =r\langle
j^{2}\rangle $ is expressed as%
\begin{eqnarray}
\langle j_{\phi }\rangle &=&-\frac{eqw^{6}}{4\pi ^{2}a^{5}}\int_{0}^{\infty
}d\lambda \int_{0}^{\infty }dp\int_{-\infty }^{+\infty }dk_{z}\sum_{j=\pm
1/2,...}\epsilon _{j}\frac{\lambda ^{2}p}{E}  \notag \\
&&\times J_{\beta _{j}}(\lambda r)J_{\beta _{j}+\epsilon _{j}}(\lambda r)%
\frac{G_{\nu _{2},\nu _{1}}^{2}(pw_{0},pw)+G_{\nu _{2},\nu
_{2}}^{2}(pw_{0},pw)}{J_{\nu _{2}}^{2}(pw_{0})+Y_{\nu _{2}}^{2}(pw_{0})},
\label{j2R}
\end{eqnarray}%
where the energy is given by (\ref{E}). If we write the parameter $\alpha $
in the form
\begin{equation}
\alpha =n_{0}+\alpha _{0}\ ,\ |\alpha _{0}|<1/2\ ,  \label{alfa0}
\end{equation}%
where $n_{0}$ is an integer number, then, by shifting $j+n_{0}\rightarrow j$%
, we can see that the VEV depends on $\alpha _{0}$ only.

For the transformation of the expression (\ref{j2R}) we use the integral
representation%
\begin{equation}
\frac{1}{E}=\frac{2}{\sqrt{\pi }}\int_{0}^{\infty }dv\ e^{-(\lambda
^{2}+p^{2}+k_{z}^{2})v^{2}}\ .  \label{ident1}
\end{equation}%
For the evaluation of the integral over $\lambda $ we use the formula%
\begin{equation}
\int_{0}^{\infty }d\lambda \,\lambda ^{2}e^{-\lambda ^{2}v^{2}}J_{\beta
_{j}}(\lambda r)J_{\beta _{j}+\epsilon _{j}}(\lambda r)=\epsilon _{j}\frac{%
re^{-u}}{4v^{4}}\left[ I_{\beta _{j}}(u)-I_{\beta _{j}+\epsilon _{j}}(u)%
\right] \ ,  \label{Int1}
\end{equation}%
with $u=r^{2}/(2v^{2})$. After evaluation of the $k_{z}$-integral and
introducing instead of $v$ a new integration variable $u=r^{2}/(2v^{2})$,
the current density is expressed as%
\begin{eqnarray}
\langle j_{\phi }\rangle &=&-\frac{eqw^{6}}{4\pi ^{2}a^{5}r^{3}}%
\int_{0}^{\infty }du\,ue^{-u}\mathcal{J}(q,\alpha _{0},u)\int_{0}^{\infty
}dp\,p  \notag \\
&&\times e^{-p^{2}r^{2}/(2u)}\frac{G_{\nu _{2},\nu
_{1}}^{2}(pw_{0},pw)+G_{\nu _{2},\nu _{2}}^{2}(pw_{0},pw)}{J_{\nu
_{2}}^{2}(pw_{0})+Y_{\nu _{2}}^{2}(pw_{0})},  \label{j2R2}
\end{eqnarray}%
where we have introduced the notation%
\begin{equation}
\mathcal{J}(q,\alpha _{0},u)=\sum_{j}\left[ I_{\beta _{j}}(u)-I_{\beta
_{j}+\epsilon _{j}}(u)\right] .  \label{seriesI0}
\end{equation}

An alternative expression is obtained from (\ref{j2R2}) by using the
integral representation for the function $\mathcal{J}(q,\alpha _{0},u)$ \cite%
{Mello2013,Moha15,Beze15}. The latter is obtained from the representation
for the series $\sum_{j}I_{\beta _{j}}(u)$ from \cite{Mello}:
\begin{eqnarray}
\mathcal{J}(q,\alpha _{0},u) &=&\frac{4}{\pi }\int_{0}^{\infty }dx\frac{%
e^{-u\cosh (2x)}g(q,\alpha _{0},2x)\cosh x}{\cosh (2qx)-\cos (q\pi )}  \notag
\\
&&+\frac{4}{q}\sideset{}{'}{\sum}_{k=1}^{[q/2]}(-1)^{k}s_{k}\sin \left( 2\pi
k\alpha _{0}\right) e^{u\cos (2\pi k/q)},  \label{Jcal2}
\end{eqnarray}%
where $s_{k}=\sin \left( \pi k/q\right) $ and
\begin{equation}
g(q,\alpha _{0},x)=\sum_{\chi =-,+}\chi \cos \left[ q\pi \left( 1/2+\chi
\alpha _{0}\right) \right] \cosh \left[ \left( 1/2-\chi \alpha _{0}\right) qx%
\right] .  \label{gqa}
\end{equation}%
The prime on the summation sign in (\ref{Jcal2}) means that for even values
of $q$ the term with $k=q/2$ should be taken with an additional coefficient
1/2. With (\ref{Jcal2}), after evaluating the $u$-integral, from (\ref{j2R2}%
) we find%
\begin{eqnarray}
\langle j_{\phi }\rangle &=&-\frac{ew^{6}}{2\pi ^{2}a^{5}r}\int_{0}^{\infty
}dp\,p^{3}\frac{G_{\nu _{2},\nu _{1}}^{2}(pw_{0},pw)+G_{\nu _{2},\nu
_{2}}^{2}(pw_{0},pw)}{J_{\nu _{2}}^{2}(pw_{0})+Y_{\nu _{2}}^{2}(pw_{0})}
\notag \\
&&\times \left[ \sideset{}{'}{\sum}_{k=1}^{[q/2]}\frac{(-1)^{k}}{s_{k}}\sin
\left( 2\pi k\alpha _{0}\right) K_{2}(2prs_{k})\right.  \notag \\
&&\left. +\frac{q}{\pi }\int_{0}^{\infty }dx\frac{g(q,\alpha _{0},2x)/\cosh x%
}{\cosh (2qx)-\cos (q\pi )}K_{2}(2pr\cosh x)\right] ,  \label{J2R3}
\end{eqnarray}%
where $K_{\nu }(x)$ is the Macdonald function.

We are interested in the effect induced by the brane. In order to separate
the corresponding contribution one should subtract from (\ref{J2R3}) the
current density in the absence of the brane. The latter is evaluated by the
formula (\ref{jmu}) with the mode functions $\psi _{\sigma }^{(\pm )}(x)$
replaced by the functions $\psi _{(0)\sigma }^{(\pm )}(x)$ in the brane-free
geometry. Those functions are obtained from (\ref{FermMod}) with the
function $W_{\nu }(pw)=J_{\nu }(pw)$ and the normalization constant
\begin{equation}
\left\vert C_{(0)\sigma }^{(\pm )}\right\vert ^{-2}=\frac{16\pi
^{2}a^{4}\eta }{q\lambda ^{2}p^{2}}E\sqrt{E^{2}-\lambda ^{2}}\kappa _{\eta
}b_{\eta }^{(\pm )}.  \label{C0}
\end{equation}%
For the VEV of the azimuthal current density in the brane-free geometry this
gives%
\begin{eqnarray}
\langle j_{\phi }\rangle _{0} &=&-\frac{eqw^{6}}{4\pi ^{2}a^{5}}%
\int_{0}^{\infty }d\lambda \int_{0}^{\infty }dp\int_{-\infty }^{+\infty
}dk_{z}\sum_{j=\pm 1/2,...}\epsilon _{j}\frac{\lambda ^{2}p}{E}  \notag \\
&&\times J_{\beta _{j}}(\lambda r)J_{\beta _{j}+\epsilon _{j}}(\lambda r)
\left[ J_{\nu _{1}}^{2}(pw)+J_{\nu _{2}}^{2}(pw)\right] .  \label{J20}
\end{eqnarray}%
By transformations similar to those we have described before, this VEV\ is
presented as
\begin{eqnarray}
\langle j_{\phi }\rangle _{0} &=&-\frac{ew^{6}}{2\pi ^{2}a^{5}r}%
\int_{0}^{\infty }dp\,p^{3}\left[ J_{\nu _{1}}^{2}(pw)+J_{\nu _{2}}^{2}(pw)%
\right]  \notag \\
&&\times \left[ \sideset{}{'}{\sum}_{k=1}^{[q/2]}\frac{(-1)^{k}}{s_{k}}\sin
\left( 2\pi k\alpha _{0}\right) K_{2}(2prs_{k})\right.  \notag \\
&&\left. +\frac{q}{\pi }\int_{0}^{\infty }dx\frac{g(q,\alpha _{0},2x)/\cosh x%
}{\cosh (2qx)-\cos (q\pi )}K_{2}(2pr\cosh x)\right] .  \label{j2R0}
\end{eqnarray}%
Note that (\ref{j2R0}) is the same for $s=+1$ and $s=-1$ and the current
densities in the brane-free geometry coincide for two inequivalent
representations of the Clifford algebra.

We present the current density in the geometry with the brane in the
decomposed form%
\begin{equation}
\langle j_{\phi }\rangle =\langle j_{\phi }\rangle _{0}+\langle j_{\phi
}\rangle _{\mathrm{b}},  \label{J2dec}
\end{equation}%
where the part $\langle j_{\phi }\rangle _{\mathrm{b}}$ is induced by the
brane. The corresponding integral representation is obtained by subtracting (%
\ref{j2R0}) from (\ref{J2R3}). We can further transform the expression for
the brane-induced contribution by making use of relation

\begin{equation}
\frac{G_{\nu ,\mu }^{2}(x,y)}{J_{\nu }^{2}(x)+Y_{\nu }^{2}(x)}-J_{\mu
}^{2}(y)=-\frac{1}{2}\sum_{n=1,2}\frac{J_{\nu }(x)}{H_{\nu }^{(n)}(x)}H_{\mu
}^{(n)2}(y),  \label{Rel1}
\end{equation}%
where $H_{\nu }^{(n)}(x)$, $n=1,2$, are the Hankel functions. This leads to
the expression%
\begin{eqnarray}
\langle j_{\phi }\rangle _{\mathrm{b}} &=&\frac{ew^{6}}{4\pi ^{2}a^{5}r}%
\sum_{l=1,2}\int_{0}^{\infty }dp\,p^{3}\sum_{n=1,2}\frac{J_{\nu _{2}}(pw_{0})%
}{H_{\nu _{2}}^{(n)}(pw_{0})}H_{\nu _{l}}^{(n)2}(pw)  \notag \\
&&\times \left[ \sideset{}{'}{\sum}_{k=1}^{[q/2]}\frac{(-1)^{k}}{s_{k}}\sin
\left( 2\pi k\alpha _{0}\right) K_{2}(2prs_{k})\right.  \notag \\
&&\left. +\frac{q}{\pi }\int_{0}^{\infty }dx\frac{g(q,\alpha _{0},2x)/\cosh x%
}{\cosh (2qx)-\cos (q\pi )}K_{2}(2pr\cosh x)\right] .  \label{J2R4}
\end{eqnarray}%
As the next step we rotate the contour of the integration over $p$ by the
angle $\pi /2$ ($-\pi /2$) for the term with $n=1$ ($n=2$). Introducing the
modified Bessel functions one finds%
\begin{eqnarray}
\langle j_{\phi }\rangle _{\mathrm{b}} &=&\frac{ew^{6}}{2\pi ^{2}a^{5}r}%
\int_{0}^{\infty }dp\,p^{3}\frac{I_{\nu _{2}}(pw_{0})}{K_{\nu _{2}}(pw_{0})}%
\left[ K_{\nu _{2}}^{2}(pw)-K_{\nu _{1}}^{2}(pw)\right]  \notag \\
&&\times \left[ \sideset{}{'}{\sum}_{k=1}^{[q/2]}\frac{(-1)^{k}}{s_{k}}\sin
\left( 2\pi k\alpha _{0}\right) J_{2}(2prs_{k})\right.  \notag \\
&&\left. +\frac{q}{\pi }\int_{0}^{\infty }dx\frac{g(q,\alpha _{0},2x)/\cosh x%
}{\cosh (2qx)-\cos (q\pi )}J_{2}(2pr\cosh x)\right] .  \label{J2R5}
\end{eqnarray}%
From (\ref{J2R5}) it follows that the current density is an odd function of $%
\alpha _{0}$. For a massless field one has $\nu _{1}=-s/2$, $\nu _{2}=s/2$
and the brane-induced contribution in the R-region vanishes. The charge flux
through the hypersurface $\phi =\mathrm{const}$ is expressed as $n_{\phi
}\langle j_{\phi }\rangle _{\mathrm{b}}$, where $n_{\phi }=a/w$ is the
normal to the hypersurface. Note that the quantity $n_{\phi }\langle j_{\phi
}\rangle _{\mathrm{b}}$ depends on $r,w,w_{0}$ through the ratios $r/w$ and $%
w_{0}/w$. The first one is the proper distance from the string
\begin{equation}
r_{\mathrm{p}}=ar/w,  \label{rp}
\end{equation}%
measured in units of the AdS curvature radius, and the second one determines
the proper distance from the brane%
\begin{equation}
y-y_{0}=a\ln (w/w_{0}).  \label{DistProp}
\end{equation}

For half-integer values of the ratio of the magnetic flux to the flux
quantum one finds%
\begin{equation}
\lim_{\alpha _{0}\rightarrow \pm 1/2}\langle j_{\phi }\rangle _{\mathrm{b}%
}=\mp \frac{eqw^{6}}{2\pi ^{3}a^{5}r}\int_{0}^{\infty }dp\,p^{3}\frac{I_{\nu
_{2}}(pw_{0})}{K_{\nu _{2}}(pw_{0})}\left[ K_{\nu _{2}}^{2}(pw)-K_{\nu
_{1}}^{2}(pw)\right] \int_{1}^{\infty }du\,\frac{J_{2}(2pru)}{u\sqrt{u^{2}-1}%
}.  \label{J2Rhi}
\end{equation}%
This shows that the brane-induced contribution is discontinuous at half
integer values of the magnetic flux in units of the flux quantum. The same
is the case for the brane-free contribution. It is of interest to note that
the limiting values $\lim_{\alpha _{0}\rightarrow \pm 1/2}\langle j_{\phi
}\rangle _{\mathrm{b}}$ are linear functions of the parameter $q$.

For $q=1$ the planar angle deficit is absent and from (\ref{J2R4}) we get
\begin{eqnarray}
\langle j_{\phi }\rangle _{\mathrm{b}} &=&-\frac{e\sin \left( \pi \alpha
_{0}\right) w^{6}}{4\pi ^{3}a^{5}r}\int_{0}^{\infty }dp\,p^{3}\frac{I_{\nu
_{2}}(pw_{0})}{K_{\nu _{2}}(pw_{0})}\left[ K_{\nu _{2}}^{2}(pw)-K_{\nu
_{1}}^{2}(pw)\right]  \notag \\
&&\times \int_{0}^{\infty }dx\frac{\cosh (2\alpha _{0}x)}{\cosh (2x)+1}%
J_{2}(2pr\cosh x).  \label{J2R5q1}
\end{eqnarray}%
This expression describes the brane-induced contribution in the current
density generated by an idealized zero-thickness magnetic flux tube. We
expect that it will approximate the corresponding current density for a
finite thickness flux tube at distances from the tube larger than the core
radius.

Note that the integral in the expression (\ref{j2R0}) for the current
density $\langle j_{\phi }\rangle _{0}$ is evaluated by using the formula
\cite{Prud2} (in \cite{Prud2} there is a missprint, instead of $%
(u^{2}-1)^{-(1\mp 2\nu )/4}$ should be $(u^{2}-1)^{-(1\pm 2\nu )/4}$)%
\begin{equation}
\int_{0}^{\infty }dp\,p^{3}J_{\nu _{l}}^{2}(pw)K_{2}(pc)=\frac{c^{2}e^{-5\pi
i/2}Q_{\nu _{l}-1/2}^{5/2}(u)}{\sqrt{2\pi }w^{6}\left( u^{2}-1\right) ^{5/4}}%
,  \label{Int2}
\end{equation}%
where $u=1+c^{2}/(2w^{2})$ and $Q_{\nu }^{\mu }(u)$ is the associated
Legendre function of the second kind. This gives (see also \cite{Oliveira})%
\begin{eqnarray}
\langle j_{\phi }\rangle _{0} &=&-\frac{\sqrt{2}er}{\pi ^{5/2}a^{5}}\left[ %
\sideset{}{'}{\sum}_{k=1}^{[q/2]}(-1)^{k}s_{k}\sin \left( 2\pi k\alpha
_{0}\right) \mathcal{Z}_{ma}(1+2\left( rs_{k}/w\right) ^{2})\right.  \notag
\\
&&\left. +\frac{q}{\pi }\int_{0}^{\infty }dx\frac{g(q,\alpha _{0},2x)\cosh x%
}{\cosh (2qx)-\cos (q\pi )}\mathcal{Z}_{ma}(1+2\left( r\cosh x/w\right) ^{2})%
\right] ,  \label{j20}
\end{eqnarray}%
with the notation%
\begin{equation}
\mathcal{Z}_{ma}(u)=e^{-5i\pi /2}\frac{Q_{ma}^{5/2}\left( u\right)
+Q_{ma-1}^{5/2}\left( u\right) }{\left( u^{2}-1\right) ^{5/4}}.  \label{Zu}
\end{equation}%
Near the string the leading term in the expansion over the distance from the
string is given by
\begin{equation}
\langle j_{\varphi }\rangle _{0}\approx -\frac{3e(w/a)^{5}}{16\pi ^{2}r^{4}}%
h_{4}(q,\alpha _{0}),  \label{j20near}
\end{equation}%
where we have introduced the notation%
\begin{equation}
h_{\mu }(q,\alpha _{0})=\sideset{}{'}{\sum}_{k=1}^{[q/2]}(-1)^{k}s_{k}^{-\mu
}\sin \left( 2\pi k\alpha _{0}\right) +\frac{q}{\pi }\int_{0}^{\infty }dx%
\frac{g(q,\alpha _{0},2x)\cosh ^{-\mu }x}{\cosh (2qx)-\cos (q\pi )}.
\label{hmu}
\end{equation}%
At large distances from the string, $r\gg w$, the leading term in the
corresponding expansion is expressed as%
\begin{equation}
\langle j_{\phi }\rangle _{0}\approx -e\frac{(2ma+3)(2ma+1)w}{\pi
^{2}a^{5}\left( 2r/w\right) ^{2ma+4}}h_{2ma+4}(q,\alpha _{0}).
\label{j20large}
\end{equation}%
For a massless field the expression of the current density is simplified to%
\begin{equation}
\langle j_{\phi }\rangle _{0}=-\frac{3e(w/a)^{5}}{16\pi ^{2}r^{4}}%
h_{4}(q,\alpha _{0}).  \label{j20m0}
\end{equation}%
Note that for a massive field the leading term in the expansion over the
distance from the string, given by (\ref{j20near}), coincides with (\ref%
{j20m0}).

The VEV of the current density for cosmic string in the Minkowski bulk, $%
\langle j_{\phi }\rangle _{0}^{\mathrm{(M)}}$, is obtained from $\langle
j_{\phi }\rangle _{0}$ in the limit $a\rightarrow \infty $ with fixed value
of the coordinate $y$. In this limit one has $w\approx a+y$ and for the
Bessel functions in (\ref{j2R0}) we can use the Debye's asymptotic
expansions \cite{Abra}. This leads to the expression%
\begin{eqnarray}
\langle j_{\phi }\rangle _{0}^{\mathrm{(M)}} &=&-\frac{e}{\pi ^{3}r}%
\int_{m}^{\infty }dp\,\frac{p^{3}}{\sqrt{p^{2}-m^{2}}}\left[ %
\sideset{}{'}{\sum}_{k=1}^{[q/2]}\frac{(-1)^{k}}{s_{k}}\sin \left( 2\pi
k\alpha _{0}\right) K_{2}(2prs_{k})\right.  \notag \\
&&\left. +\frac{q}{\pi }\int_{0}^{\infty }dx\frac{g(q,\alpha _{0},2x)}{\cosh
(2qx)-\cos (q\pi )}\frac{K_{2}(2pr\cosh x)}{\cosh x}\right] .  \label{j0M}
\end{eqnarray}%
The integrals are evaluated by using the formula from \cite{Prud2} and we
find%
\begin{eqnarray}
\langle j_{\phi }\rangle _{0}^{\mathrm{(M)}} &=&-\frac{em^{5/2}}{2\pi
^{5/2}r^{3/2}}\left[ \sideset{}{'}{\sum}_{k=1}^{[q/2]}\frac{(-1)^{k}}{%
s_{k}^{3/2}}\sin \left( 2\pi k\alpha _{0}\right) K_{5/2}\left(
2mrs_{k}\right) \right.  \notag \\
&&\left. +\frac{q}{\pi }\int_{0}^{\infty }dx\frac{g(q,\alpha _{0},2x)}{\cosh
(2qx)-\cos (q\pi )}\frac{K_{5/2}(2mr\cosh x)}{\cosh ^{3/2}x}\right] .
\label{j0M2}
\end{eqnarray}%
For a massless field one gets $\langle j_{\phi }\rangle _{0}^{\mathrm{(M)}%
}=-3eh_{4}(q,\alpha _{0})/(16\pi ^{2}r^{4})$. This result is conformally
related to the expression (\ref{j20m0}) of the current density for cosmic
string in the AdS bulk. For a massive field, at large distances from the
string, $mr\gg 1$, the current density $\langle j_{\phi }\rangle _{0}^{%
\mathrm{(M)}}$ is suppresses by the factor $\exp [-2mr\sin (\pi /q)]$ for $%
q\geq 2$ and by the factor $e^{-2mr}$ for $1\leq q<2$. Recall that for the
AdS bulk and at large distances from the string, the brane-free contribution
in the current density decays according to inverse power-law (see (\ref%
{j20large})).

\begin{figure}[tbph]
\begin{center}
\epsfig{figure=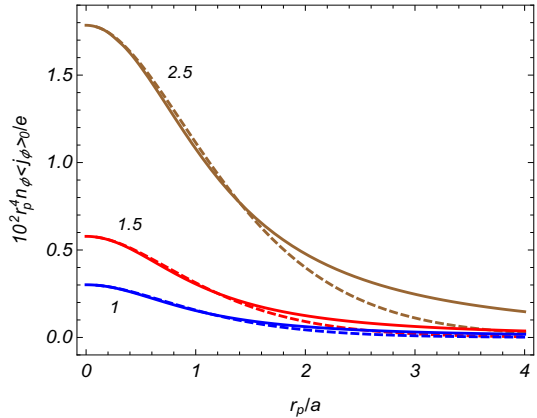,width=7.cm,height=5.5cm}
\end{center}
\caption{The boundary-free parts in the current density for the AdS (full
curves) and Minkowski (dashed curves) background geometries as functions of
the proper distance from the string. The graphs are plotted for $ma=1$, $%
\protect\alpha _{0}=0.3$ and the numbers near the curves are the
corresponding values of the parameter $q$. }
\label{fig1}
\end{figure}

The full curves in figure \ref{fig1} present the quantity $r_{\mathrm{p}%
}^{4}n_{\phi }\langle j_{\phi }\rangle _{0}/e$ for cosmic string in the AdS
bulk as a function of the proper distance from the string (\ref{rp}),
measured in units of the AdS curvature radius. The graphs are plotted for $%
ma=1$, $\alpha _{0}=0.3$, and the numbers near the curves correspond to the
values of the parameter $q$. The same quantity for cosmic string in the
Minkowski is plotted in figure \ref{fig1} by dashed curves. The
corresponding line element is given by the expression in the brackets on the
right-hand side of (\ref{ds2}) and $r_{\mathrm{p}}=r$, $n_{\phi }=1$. For
both background geometries $r_{\mathrm{p}}^{4}n_{\phi }\langle j_{\phi
}\rangle _{0}/e|_{r=0}=-3h_{4}(q,\alpha _{0})/(16\pi ^{2})$ and this
limiting value is equal to $0.00301$, $0.00578$, $0.0179$ for $q=1,1.5,2.5$,
respectively. As noted above, near the string the leading terms in the
corresponding asymptotic expansions for the AdS and Minkowski bulks coincide
and in that region the effects of the background curvature are weak. As seen
from the graphs and in accordance with the asymptotic analysis, those
effects are essential at large distances from the cosmic string. Note that
for a massless field one has $r_{\mathrm{p}}^{4}n_{\phi }\langle j_{\phi
}\rangle _{0}/e=-3h_{4}(q,\alpha _{0})/(16\pi ^{2})$ (see (\ref{j20m0})) for
all values of $r_{\mathrm{p}}$ in both AdS and Minkowski geometries. For $%
q=1,1.5,2.5$ the corresponding graphs will be presented by horizontal lines
which touch the curves in figure \ref{fig1} at $r_{\mathrm{p}}=0$.

\subsection{Asymptotics of the current density and numerical results}

We turn to the investigation of the current density for the geometry with a
brane in asymptotic regions of the variables. Near the string, assuming that
$r\ll w-w_{0}$, we get%
\begin{equation}
\langle j_{\phi }\rangle _{\mathrm{b}}\approx \frac{erh_{-1}(q,\alpha _{0})}{%
4\pi ^{2}a^{5}}\int_{0}^{\infty }dx\,x^{5}\frac{I_{\nu _{2}}(xw_{0}/w)}{%
K_{\nu _{2}}(xw_{0}/w)}\left[ K_{\nu _{2}}^{2}(x)-K_{\nu _{1}}^{2}(x)\right]
,  \label{NearStr}
\end{equation}%
and the boundary-induced VEV behaves as $\langle j_{\phi }\rangle _{\mathrm{b%
}}\propto r$. Note that on the string the brane free part $\langle j_{\phi
}\rangle _{0}$ diverges and, hence, it dominates for points near the string.
For the investigation of the current density at large distances from the
string it is convenient to use the representation (\ref{J2R3}). Introducing
a new integration variable $u=pr$ and assuming that $r\gg w$, we see that
the dominant contribution comes from the part of the integrand with the
function $G_{\nu _{2},\nu _{1}}(uw_{0}/r,uw/r)$. By taking into account that%
\begin{equation}
G_{\nu _{2},\nu _{1}}(uw_{0}/r,uw/r)\approx -\frac{2sr}{\pi }\frac{%
(w/w_{0})^{s\nu _{1}}}{uw_{0}},  \label{gl}
\end{equation}%
to the leading order one finds%
\begin{equation}
\langle j_{\phi }\rangle \approx -4e|\nu _{2}|\left( |\nu _{2}|+1\right)
h_{2|\nu _{2}|+3}(q,\alpha _{0})\frac{w(w_{0}/w)^{2|\nu _{2}|-2s\nu _{2}}}{%
\pi ^{2}a^{5}(2r/w)^{2|\nu _{2}|+3}}.  \label{J2Rlarge}
\end{equation}%
As seen, the decay of the current density at large distances from the string
is power-law for both massless and massive fields. This is in contrast to
the case of the Minkowski bulk, where the decay for massive fields is
exponential. For a massless field the brane-induced contribution vanishes
and (\ref{J2Rlarge}) is reduced to (\ref{j20m0}). For $s=1$ the leading term
(\ref{J2Rlarge}) coincides with (\ref{j20large}) and, hence, the current
density at large distances is dominated by the brane-free contribution. For
the field with $s=-1$, at large distances from the string the current
density behaves as $\langle j_{\phi }\rangle \propto (w/r)^{|2ma-1|+3}$ and
the brane-induced contribution dominates.

Now let us consider the behavior of the current density at large distances
from the brane, $w\gg w_{0}$. Introducing in (\ref{J2R5}) a new integration
variable $u=pw$, we use the asymptotic expressions for the modified Bessel
functions with the argument $uw_{0}/w$ for small values of the latter ratio
\cite{Abra}. It can be seen that, to the leading order, the brane induced
current density is suppressed by the factor $(w_{0}/2w)^{|\nu _{2}|+\nu
_{2}} $.This factor also describes the behavior of the current density for
fixed $w $ when the location of the brane tends to the AdS boundary, $%
w_{0}\rightarrow 0$. If in addition $w\gg w_{0}$ one has $w\gg r$, we use
the asymptotic of the Bessel function for small argument. After integrating
over $u$ and for $\nu _{2}>0$ this gives%
\begin{equation}
\langle j_{\phi }\rangle _{\mathrm{b}}\approx \frac{er(w_{0}/2w)^{2\nu _{2}}%
}{8\pi ^{3/2}a^{5}\nu _{2}}\frac{\Gamma \left( 3+\nu _{2}\right)
h_{-1}(q,\alpha _{0})}{\Gamma ^{2}(\nu _{2})\Gamma \left( 7/2+\nu
_{2}\right) }\left[ 2\Gamma \left( 3+2\nu _{2}\right) -\Gamma \left( 3+2\nu
_{2}-s\right) \Gamma \left( 3+s\right) \right] .  \label{J2hor}
\end{equation}%
Considered as a function of the physical distance from the brane, $y-y_{0}$,
the brane-induced contribution is suppressed by the factor $e^{-2\nu
_{2}(y-y_{0})/a}$. For $\nu _{2}<0$ the leading term in the asymptotic
expansion is given by
\begin{equation}
\langle j_{\phi }\rangle _{\mathrm{b}}\approx \frac{2eh_{-1}(q,\alpha _{0})r%
}{15\pi ^{2}a^{5}}\sum_{l=1,2}\left( 4-\nu _{l}^{2}\right) \left( \nu
_{l}^{2}-1\right) \nu _{l}.  \label{J2hor2}
\end{equation}%
In this case, the brane-induced part tends to finite value on the AdS
horizon.

For the investigation of the current density near the brane it is convenient
to use the representation (\ref{J2R3}). In particular, the current density
on the brane is directly obtained putting in the integrand $w=w_{0}$. By
taking into account that $G_{\nu _{2},\nu _{1}}(x,x)=-2s/(\pi x)$, we get
\begin{eqnarray}
\langle j_{\phi }\rangle |_{w=w_{0}} &=&-\frac{2ew_{0}^{2}}{\pi ^{4}a^{5}r}%
\int_{0}^{\infty }dp\frac{\,p}{J_{\nu _{2}}^{2}(p)+Y_{\nu _{2}}^{2}(p)}
\notag \\
&&\times \left[ \sideset{}{'}{\sum}_{k=1}^{[q/2]}\frac{(-1)^{k}}{s_{k}}\sin
\left( 2\pi k\alpha _{0}\right) K_{2}\left( \frac{2pr}{w_{0}}s_{k}\right)
\right.  \notag \\
&&\left. +\frac{q}{\pi }\int_{0}^{\infty }dx\frac{g(q,\alpha _{0},2x)/\cosh x%
}{\cosh (2qx)-\cos (q\pi )}K_{2}\left( \frac{2pr}{w_{0}}\cosh x\right) %
\right] .  \label{J2RonBr}
\end{eqnarray}%
Unlike the fermion condensate and the VEV of the energy-momentum tensor, the
VEV of the current density is finite on the brane. This feature can be
understood as follows. In the absence of the magnetic flux the current
density vanishes in the geometry with the brane. In the presence of the
localized flux, in the region outside the localization the local geometry
and the field strength tensor for the external electromagnetic field are the
same as in the absence of the flux (in the problem at hand the field
strength tensor is zero). From here it follows that adding the localized
flux does not change the divergence structure in the VEVs outside the
localization range. In particular, the current density being finite in the
absence of the flux remains finite in the problem with a magnetic flux.

Finally, we consider the limit $a\rightarrow \infty $ with fixed value of
the coordinate $y$. As seen from (\ref{ds2}), in that limit the geometry
under consideration is reduced to the geometry of a cosmic string in
background of (4+1)-dimensional Minkowski spacetime. For the coordinate $w$
in the arguments of the modified Bessel function one has $w\approx a+y$. By
taking into account that $\nu _{l}\gg 1$, we see that in the limit under
consideration both the order and the argument of the modified Bessel
functions in (\ref{J2R5}) are large and we can use the corresponding uniform
asymptotic expansions. From those expansions, to the leading order, one gets%
\begin{equation}
\frac{I_{\nu _{2}}(pw_{0})}{K_{\nu _{2}}(pw_{0})}\sim \frac{1}{\pi }e^{2\nu
_{2}\rho (pw_{0}/\nu _{2})},  \label{IKas}
\end{equation}%
with $\rho (x)=\sqrt{1+x^{2}}+\ln [x/(1+\sqrt{1+x^{2}})]$, and%
\begin{equation}
K_{\nu _{1}}^{2}(pw)-K_{\nu _{2}}^{2}(pw)\sim \frac{\pi e^{-2\nu _{2}\rho
(pw/\nu _{2})}}{\nu _{2}(pw/\nu _{2})^{2}}\left( \frac{1}{\sqrt{1+(pw/\nu
_{2})^{2}}}-s\right) .  \label{K2as}
\end{equation}%
In all the terms, except the arguments of the function $\rho (x)$, we can
replace $pw/\nu _{2},pw_{0}/\nu _{2}\rightarrow p/m$. Expanding the
functions in the exponents we get $\langle j_{\phi }\rangle _{\mathrm{b}%
}\rightarrow \langle j_{\phi }\rangle _{\mathrm{b}}^{\mathrm{(M)}}$, where
the current density in the Minkowskian bulk is given by
\begin{eqnarray}
\langle j_{\phi }\rangle _{\mathrm{b}}^{\mathrm{(M)}} &=&\frac{em}{2\pi ^{2}r%
}\int_{m}^{\infty }du\,e^{-2u\left( y-y_{0}\right) }\left( su-m\right)
\notag \\
&&\times \left[ \sideset{}{'}{\sum}_{k=1}^{[q/2]}\frac{(-1)^{k}}{s_{k}}\sin
\left( 2\pi k\alpha _{0}\right) J_{2}(2rs_{k}\sqrt{u^{2}-m^{2}})\right.
\notag \\
&&\left. +\frac{q}{\pi }\int_{0}^{\infty }dx\frac{g(q,\alpha _{0},2x)/\cosh x%
}{\cosh (2qx)-\cos (q\pi )}J_{2}(2r\cosh x\sqrt{u^{2}-m^{2}})\right] ,
\label{j2M}
\end{eqnarray}%
where $y=y_{0}$ specifies the location of the planar boundary in the
Minkowski bulk. For a massless field the corresponding boundary-induced
contribution vanishes. Note that the fermionic condensate, the charge and
current densities in (2+1)-dimensional conical spacetime with a single and
two circular boundaries have been investigated in \cite%
{Mello,Bell16c,Saha19,Bell20b}. The geometry of a planar ring has been
considered in \cite{Bell16b}.

\begin{figure}[tbph]
\begin{center}
\begin{tabular}{cc}
\epsfig{figure=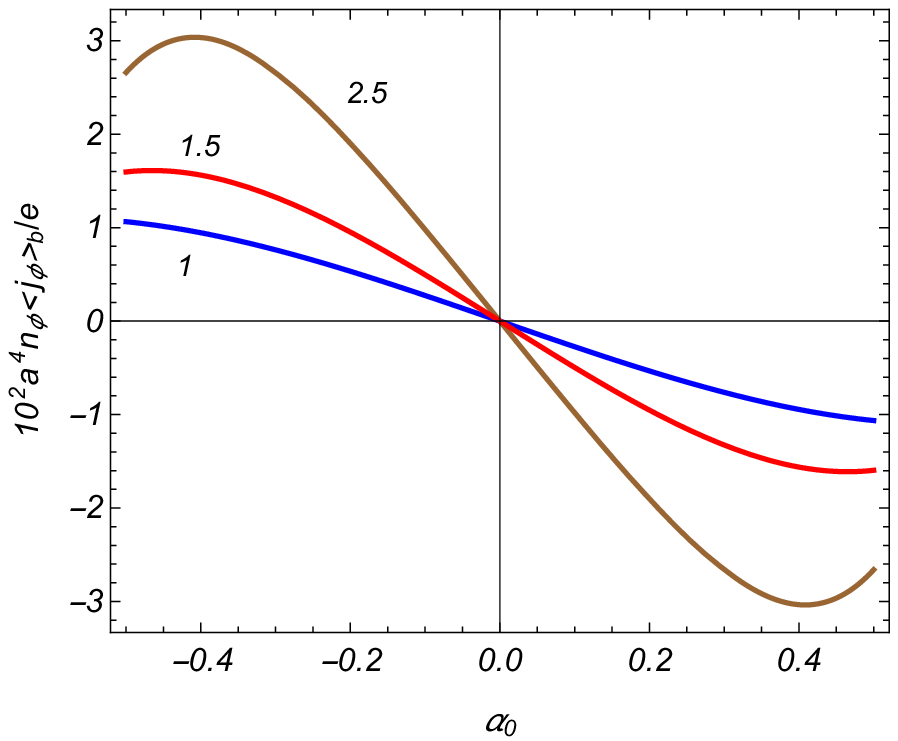,width=7.cm,height=5.5cm} & \quad %
\epsfig{figure=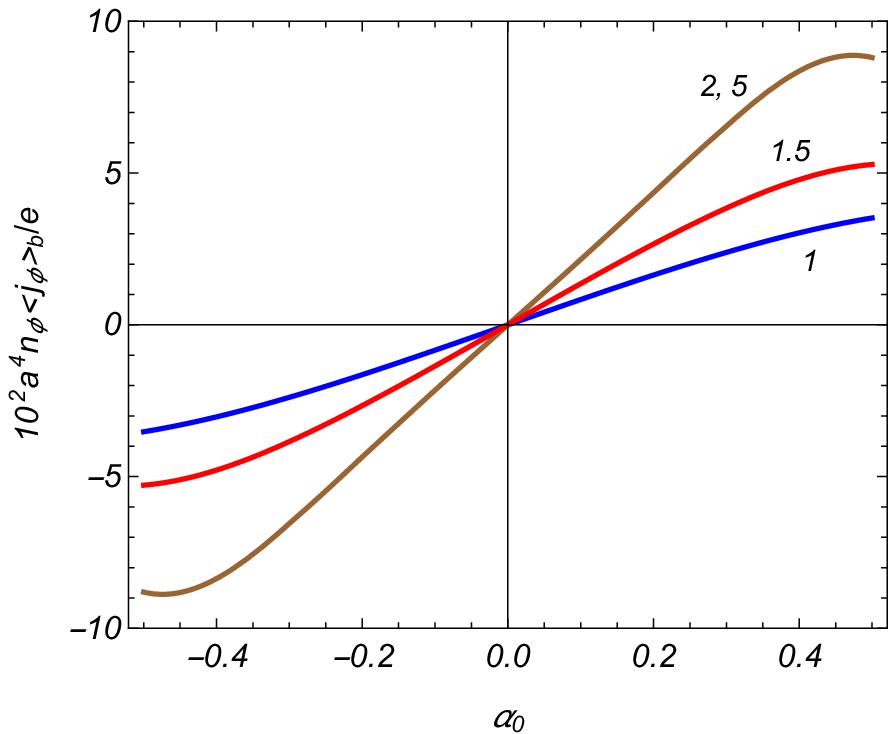,width=7.cm,height=5.5cm}%
\end{tabular}%
\end{center}
\caption{The brane-induced contribution in the VEV of the azimuthal current
density as a function of the fractional part of the ratio of the magnetic
flux to flux quantum. The left and right panels correspond to the cases $s=1$
and $s=-1$. The numbers near the curves are the values of $q$. Graphs are
plotted for $ma=1$, $w/w_{0}=1.5$, $r/w_{0}=0.5$. }
\label{fig2}
\end{figure}

In figures below we plot the charge flux through the hypersurface $\phi =%
\mathrm{const}$, given by $n_{\phi }\langle j_{\phi }\rangle _{\mathrm{b}}$,
measured in units of $e/a^{4}$. The numbers near the curves will correspond
to the values of the parameter $q$. Figure \ref{fig2} presents the
corresponding brane-induced contribution in the R-region versus the
parameter $\alpha _{0}$ that determines the fractional part of the ratio of
the magnetic flux to flux quantum. The left and right panels correspond to
fermionic fields with $s=1$ and $s=-1$, respectively. The graphs are plotted
for $ma=1$, $w/w_{0}=1.5$, $r/w_{0}=0.5$. As seen, the absolute value of the
current density increases with increasing planar angle deficit. As it has
been already mentioned, for $\alpha _{0}=\pm 1/2$ it is a linear function of
the parameter $q$. We also see that the absolute value of the brane-induced
current density for the case $s=-1$ is significantly larger compared to that
for the field with $s=1$.

\begin{figure}[tbph]
\begin{center}
\begin{tabular}{cc}
\epsfig{figure=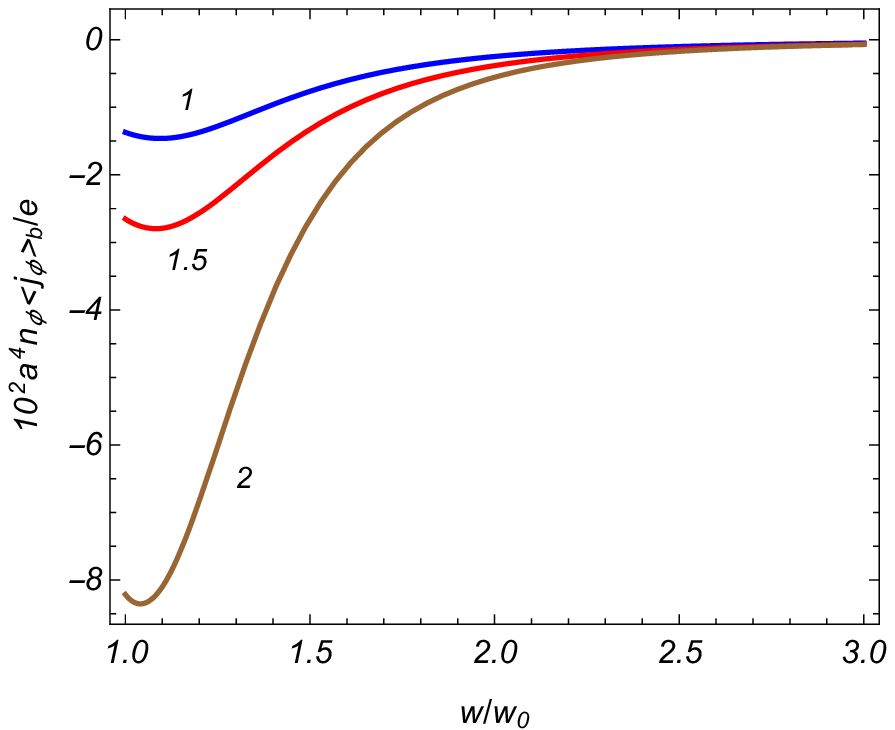,width=7.cm,height=5.5cm} & \quad %
\epsfig{figure=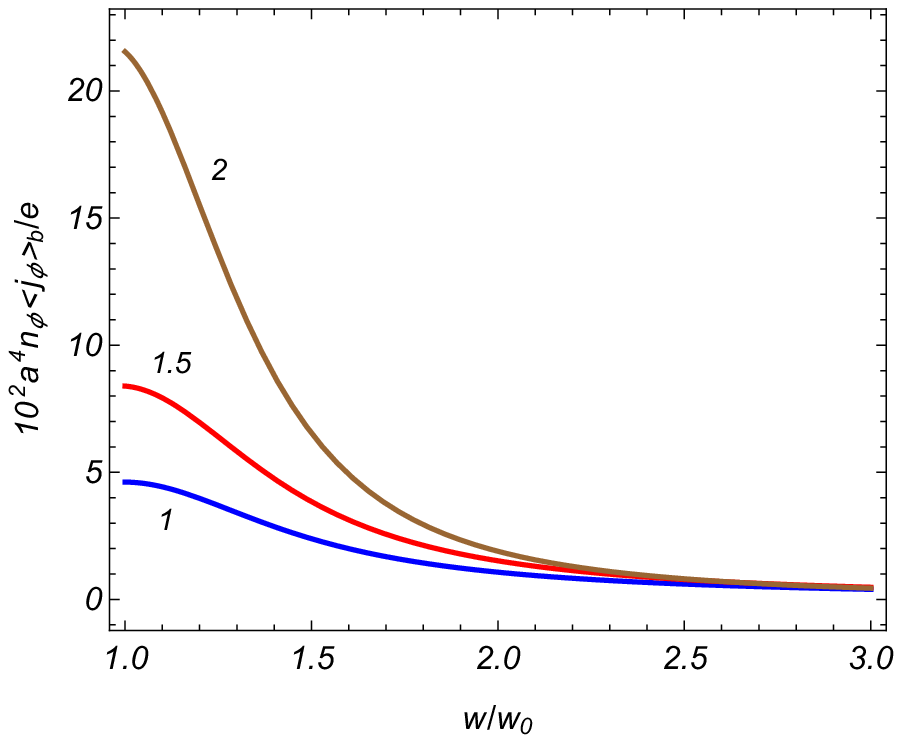,width=7.cm,height=5.5cm}%
\end{tabular}%
\end{center}
\caption{The current density induced by the brane as a function of the ratio
$w/w_{0}$ for fixed values $ma=1$, $\protect\alpha _{0}=0.3$, $r/w_{0}=0.5$.
The left and right panels correspond to fields with $s=1$ and $s=-1$.}
\label{fig3}
\end{figure}

In figure \ref{fig3} we display the dependence of the brane-induced current
density on the ratio $w/w_{0}$ for fields with $s=1$ (left panel) and $s=-1$
(right panel). For the parameters we have taken $ma=1$, $\alpha _{0}=0.3$, $%
r/w_{0}=0.5$. Note that the proper distance from the brane $y-y_{0}$ is
expressed as $y-y_{0}=a\ln (w/w_{0})$. As it has been shown by the
asymptotic analysis, for $w/w_{0}\gg 1$ the brane-induced contribution
decays as $(w/w_{0})^{2ma+s}$. The decay is stronger for the field with $s=1$%
.

\begin{figure}[tbph]
\begin{center}
\begin{tabular}{cc}
\epsfig{figure=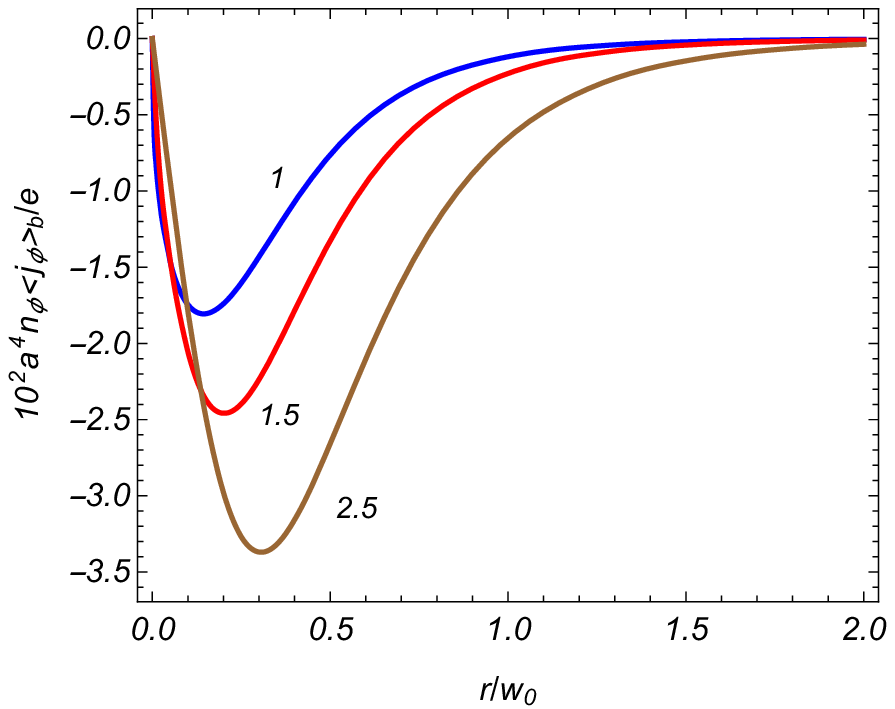,width=7.cm,height=5.5cm} & \quad %
\epsfig{figure=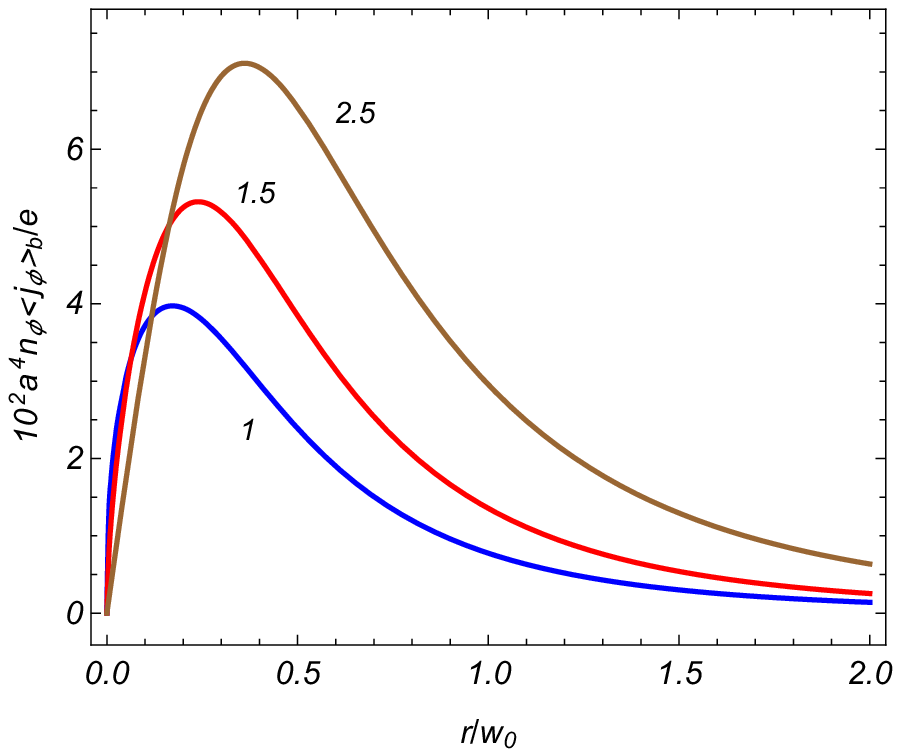,width=7.cm,height=5.5cm}%
\end{tabular}%
\end{center}
\caption{The brane-induced contribution in the VEV of the current density in
the R-region as a function of the distance from the string for fixed $ma=1$,
$\protect\alpha _{0}=0.3$, $w/w_{0}=1.5$. The left and right panels
correspond to $s=1$ and $s=-1$.}
\label{fig4}
\end{figure}

The dependence of the brane-induced current density on the distance from the
string is plotted in figure \ref{fig4} for fixed values of the combinations $%
ma=1$, $\alpha _{0}=0.3$, $w/w_{0}=1.5$. As before the left and right panels
correspond to $s=1$ and $s=-1$. The contribution of the brane linearly
vanishes on the string (see (\ref{NearStr})). At large distances from the
string it behaves like $(r/w)^{2ma+3+s}$. The fall-off is stronger for the
case $s=1$.

\begin{figure}[tbph]
\begin{center}
\begin{tabular}{cc}
\epsfig{figure=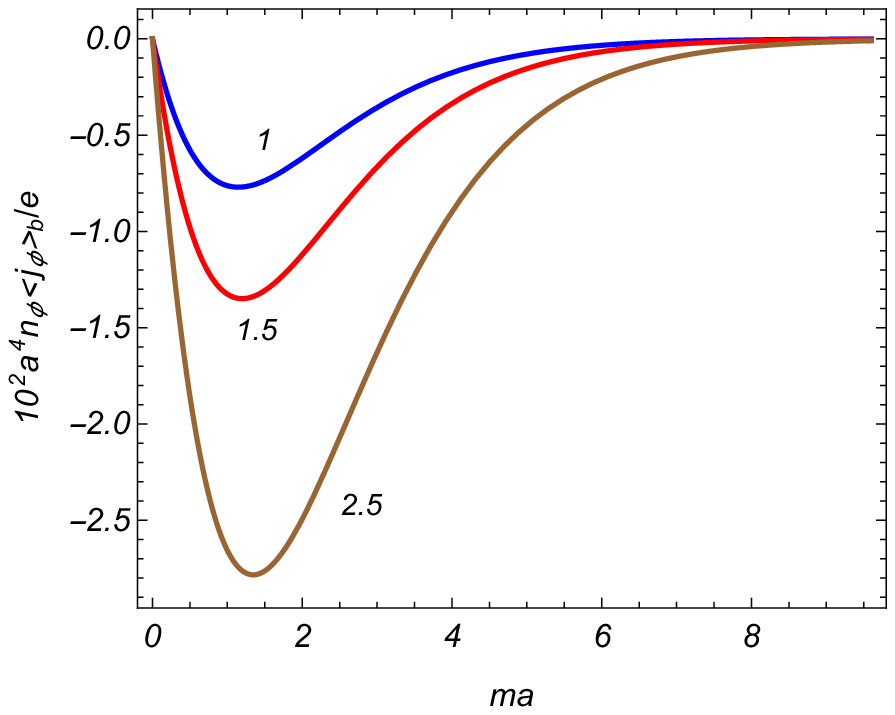,width=7.cm,height=5.5cm} & \quad %
\epsfig{figure=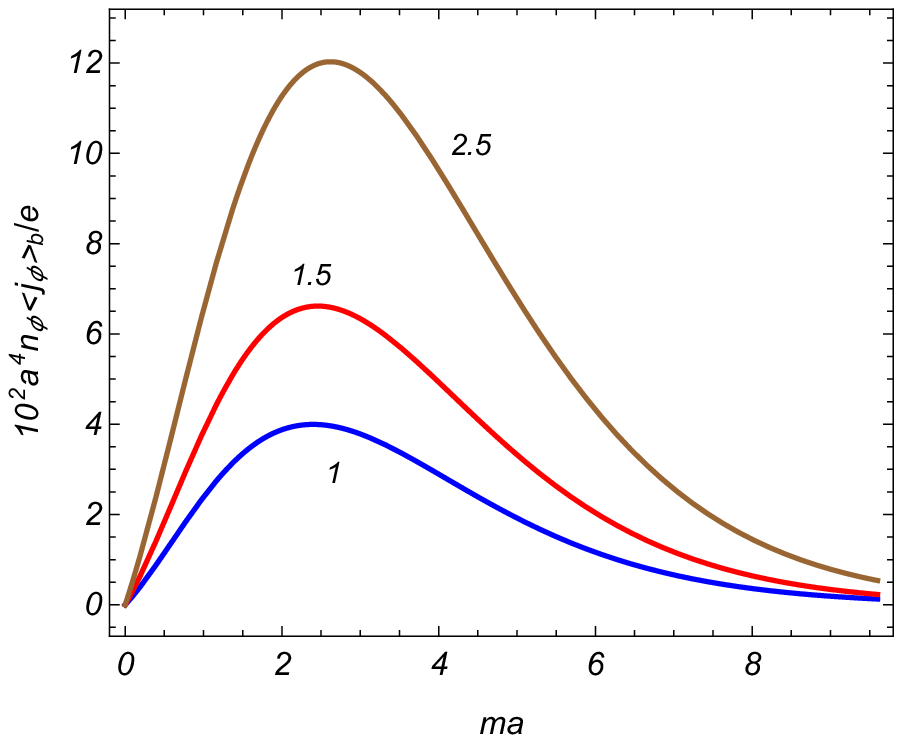,width=7.cm,height=5.5cm}%
\end{tabular}%
\end{center}
\caption{The dependence of the brane-induced current on the field mass for
fixed values $\protect\alpha _{0}=0.3$, $w/w_{0}=1.5$, $r/w_{0}=0.5$ and for
fields with $s=1$ and $s=-1$ (the left and right panels, respectively).}
\label{fig5}
\end{figure}

It is of interest to consider the dependence of the current density on the
mass of the field. Figure \ref{fig5} displays that dependence for the
brane-induced contribution for $\alpha _{0}=0.3$, $w/w_{0}=1.5$, $%
r/w_{0}=0.5 $ and for fields with $s=1$ (left panel) and $s=-1$ (right
panel). As it has been mentioned above, the brane-induced VEV in the
R-region vanishes for a massless field. We also expect that for large masses
the VEV should tend to zero and, hence, for some intermediate value of the
mass the absolute value of the current density has a maximum. That is seen
from the graphs.

\begin{figure}[tbph]
\begin{center}
\epsfig{figure=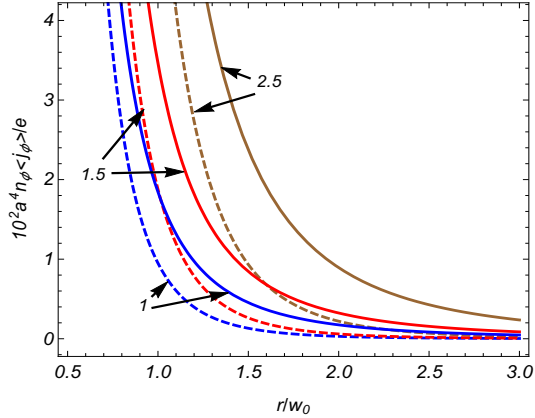,width=7.cm,height=5.5cm}
\end{center}
\caption{The total current density as a function of the radial coordinate
for fields with $s=1$ (dashed curves) and $s=-1$ (full curves). The graphs
are plotted for $ma=1$, $\protect\alpha _{0}=0.3$, $w/w_{0}=1.5$.}
\label{fig6}
\end{figure}

In the numerical examples above we have considered the brane-induced part in
the current density. To show its relative contribution, in figure \ref{fig6}
the total current density in the R-region is plotted versus the distance
from the string for fixed values $ma=1$, $\alpha _{0}=0.3$, $w/w_{0}=1.5$.
For these values the brane-free contribution is positive and monotonically
decays with increasing $r$. For the field with $s=1$ (dashed curves), though
the brane-induced part is negative, the brane-free contribution dominates at
all distances and the total current density is positive. For the field with $%
s=-1$ (full curves) both contributions are positive. For $s=-1$, the total
current density near the string is dominated by the brane-free part and at
large distances the brane-induced part dominates.

\section{Vacuum current in the L-region}

\label{sec:Lreg}

In this section we consider the VEV of the current density in the region
between the AdS boundary and the brane, corresponding to $0\leq w\leq w_{0}$.

\subsection{Integral representation}

The complete set of fermionic modes has the form (\ref{FermMod}) with the
function $W_{\nu }(pw)$ given by (\ref{Wnu}). Now, in the normalization
condition (\ref{nc}) the integration goes over the region $w\in \lbrack
0,w_{0}]$. For $ma\geq 1/2$ and $c_{2}\neq 0$ the normalization integral
diverges at the lower limit and the corresponding mode functions are not
normalizable. Hence, in this case the normalizability condition implies that
$c_{2}=0$ and $W_{\nu }(pw)=J_{\nu }(pw)$ (the constant $c_{1}$ is included
in the normalization coefficient). For $0\leq ma<1/2$ the modes with $%
c_{2}\neq 0$ are normalizable and an additional boundary condition on the
AdS boundary is required to uniquely define the wave-functions. Here we will
assume a special boundary condition corresponding to $c_{2}=0$. Hence, in
our consideration the fermionic mode functions are given by (\ref{FermMod})
with $W_{\nu }(pw)=J_{\nu }(pw)$ for $ma\geq 0$.

From the boundary condition (\ref{MITbc}) it follows that the eigenvalues
for the quantum number $p$ are roots of the equation
\begin{equation}
J_{\nu _{1}}(pw_{0})=0\ ,  \label{roots}
\end{equation}%
for both the cases $s=\pm 1$. We will denote the corresponding positive
roots with respect to $pw_{0}$ by $p_{i}=p_{i}(\nu _{1})=pw_{0}$, with $%
i=1,2,...$, assuming that they are enumerated in the ascending order, $%
p_{i+1}>p_{i}$. Note that the roots $p_{i}$ do not depend on the location of
the brane. Substituting the mode functions in (\ref{nc}) and using the
corresponding standard integral involving the products of the Bessel
functions, for the normalization constants in the L-region, $C_{\sigma
}^{(\pm )}=C_{\mathrm{(L)}\sigma }^{(\pm )}$, we get
\begin{equation}
|C_{\mathrm{(L)}\sigma }^{(\pm )}|^{2}=\frac{\eta q\lambda ^{2}p_{i}J_{\nu
_{2}}^{-2}(p_{i})}{8\pi ^{2}a^{4}E\kappa _{\eta }b_{\eta }^{(\pm )}w_{0}^{2}%
\sqrt{p_{i}^{2}+w_{0}^{2}k_{z}^{2}}}\ ,  \label{CL}
\end{equation}%
where the energy is given by $E=\sqrt{\lambda
^{2}+p_{i}^{2}/w_{0}^{2}+k_{z}^{2}}$. As seen, the normalization
coefficients are independent of the representation adopted, $s=1$ or $s=-1$.

Having specified the complete set of modes we can start the investigation of
the current density by using the formula (\ref{jmu}). Similar to the
R-region, it can be seen that the only nonzero component corresponds to the
one along the azimuthal direction. After the summation over $\eta $, we get
\begin{eqnarray}
\langle j_{\phi }\rangle &=&-\frac{eqw^{6}}{2\pi ^{2}a^{5}w_{0}^{2}r}%
\int_{0}^{\infty }d\lambda \int_{-\infty }^{+\infty
}dk_{z}\sum_{j}\sum_{i=1}^{\infty }\epsilon _{j}\frac{\lambda ^{2}}{E}
\notag \\
&&\times \frac{J_{\beta _{j}}(\lambda r)J_{\beta _{j}+\epsilon _{j}}(\lambda
r)}{J_{\nu _{2}}^{2}\left( p_{i}\right) }\left[ J_{\nu
_{1}}^{2}(p_{i}w/w_{0})+J_{\nu _{2}}^{2}(p_{i}w/w_{0})\right] \ .
\label{j2L}
\end{eqnarray}%
By using the integral representation (\ref{ident1}) and making
transformations similar to those we have described in the previous section
for the R-region, the following expression is obtained
\begin{eqnarray}
\langle j_{\phi }\rangle &=&-\frac{ew^{6}}{\pi ^{2}a^{5}rw_{0}^{4}}%
\sum_{i=1}^{\infty }p_{i}^{2}\frac{J_{\nu _{1}}^{2}(pw)+J_{\nu _{2}}^{2}(pw)%
}{J_{\nu _{2}}^{2}(p_{i})}  \notag \\
&&\times \left[ \sideset{}{'}{\sum}_{k=1}^{[q/2]}(-1)^{k}\frac{\sin \left(
2\pi k\alpha _{0}\right) }{s_{k}}K_{2}\left( 2rps_{k}\right) \right.  \notag
\\
&&\left. +\frac{q}{\pi }\int_{0}^{\infty }dx\frac{g(q,\alpha _{0},2x)/\cosh x%
}{\cosh (2qx)-\cos (q\pi )}K_{2}\left( 2rp\cosh x\right) \right] \ ,
\label{j2L2}
\end{eqnarray}%
where $p=p_{i}/w_{0}$.

In order to obtain a representation more convenient for the asymptotic and
numerical analysis, and for explicit extraction of the brane-induced
contribution, we apply to the series over $i$ a variant of the generalized
Abel-Plana formula \cite{Aram:2007},
\begin{eqnarray}
\sum_{i=1}^{\infty }\frac{f(p_{i})}{p_{i}J_{\nu _{2}}^{2}(p_{i})} &=&\frac{1%
}{2}\int_{0}^{\infty }duf(u)-\frac{1}{2\pi }\int_{0}^{\infty }du\frac{K_{\nu
_{1}}(u)}{I_{\nu _{1}}(u)}  \notag \\
&&\times \lbrack e^{-\nu _{1}\pi i}f(e^{\pi i/2}u)+e^{\nu _{1}\pi
i}f(e^{-\pi i/2}u)]\ ,  \label{AP}
\end{eqnarray}%
valid for a function $f(u)$ analytic in the right half-plane of the complex
variable $u$ (function $f(u)$ may have branch points on the imaginary axis,
for the conditions imposed on this function see \cite{Aram:2007}). For the
problem under consideration the function $f(u)$ is specified as%
\begin{equation}
f(u)=u^{3}[J_{\nu _{1}}^{2}(uw/w_{0})+J_{\nu
_{2}}^{2}(uw/w_{0})]K_{2}(2ru\gamma /w_{0}),  \label{fu}
\end{equation}%
with $\gamma =s_{k},\cosh x$. For this function one has
\begin{equation}
e^{-\nu _{1}\pi i}f(e^{\pi i/2}u)+e^{\nu _{1}\pi i}f(e^{-\pi i/2}u)=\pi
u^{3}[I_{\nu _{1}}^{2}(uw/w_{0})-I_{\nu _{2}}^{2}(uw/w_{0})]J_{2}(2ru\gamma
/w_{0}).  \label{fu2}
\end{equation}%
The contribution to the current density coming from the first term in the
right-hand side of (\ref{AP}) coincides with the current density in the
geometry without the brane given by (\ref{j2R0}). As a result, the current
density in the L-region is decomposed as (\ref{J2dec}), where the
brane-induced contribution, coming from the last term in (\ref{AP}), is
presented as
\begin{eqnarray}
\langle j_{\phi }\rangle _{\mathrm{b}} &=&\frac{ew^{6}}{2\pi ^{2}a^{5}r}%
\int_{0}^{\infty }dp\,p^{3}\frac{K_{\nu _{1}}(pw_{0})}{I_{\nu _{1}}(pw_{0})}%
\left[ I_{\nu _{1}}^{2}(pw)-I_{\nu _{2}}^{2}(pw)\right]  \notag \\
&&\times \left[ \sideset{}{'}{\sum}_{k=1}^{[q/2]}\frac{(-1)^{k}}{s_{k}}\sin
\left( 2\pi k\alpha _{0}\right) J_{2}(2prs_{k})\right.  \notag \\
&&\left. +\frac{q}{\pi }\int_{0}^{\infty }dx\frac{g(q,\alpha _{0},2x)/\cosh x%
}{\cosh (2qx)-\cos (q\pi )}J_{2}(2pr\cosh x)\right] .  \label{J2L5}
\end{eqnarray}%
It is an odd periodic function of the magnetic flux with the period of the
flux quantum.

Comparing with (\ref{J2R5}), we see that the brane-induced contribution in
the L-region is obtained from the corresponding quantity for the R-region by
the replacements $I\rightarrow K$, $K\rightarrow I$ of the modified Bessel
functions and the replacements $\nu _{1,2}\rightarrow \nu _{2,1}$ of the
orders. The expression for the brane-induced contribution in the L-region at
half-integer values of the ratio of the magnetic flux to flux quantum is
obtained from (\ref{J2Rhi}) by the same replacements. The limiting values $%
\lim_{\alpha _{0}\rightarrow \pm 1/2}\langle j_{\phi }\rangle _{\mathrm{b}}$
are linear functions of the parameter $q$ describing the planar angle
deficit. Another special case with $q=1$ corresponds to magnetic flux tube
in AdS spacetime. The corresponding expression reads
\begin{eqnarray}
\langle j_{\phi }\rangle _{\mathrm{b}} &=&\frac{e\sin \left( \pi \alpha
_{0}\right) w^{6}}{4\pi ^{3}a^{5}r}\int_{0}^{\infty }dp\,p^{3}\frac{K_{\nu
_{1}}(pw_{0})}{I_{\nu _{1}}(pw_{0})}\left[ I_{\nu _{2}}^{2}(pw)-I_{\nu
_{1}}^{2}(pw)\right]  \notag \\
&&\times \int_{0}^{\infty }dx\,\frac{\cosh (2\alpha _{0}x)}{\cosh (2x)+1}%
J_{2}(2pr\cosh x).  \label{J2L5q1}
\end{eqnarray}

For a massless field one has $K_{\nu _{1}}(x)/I_{\nu _{1}}(x)=\pi
/(e^{2x}+s) $ and $I_{\nu _{2}}^{2}(x)-I_{\nu _{1}}^{2}(x)=-2s/(\pi x)$.
Expanding $1/(e^{2x}+s)$ and evaluating the integral over $p$ by using the
result from \cite{Prud2}, from (\ref{J2L5}) we find
\begin{eqnarray}
\langle j_{\phi }\rangle _{\mathrm{b}} &=&-\frac{3er}{8\pi ^{2}(aw_{0}/w)^{5}%
}\sum_{l=1}^{\infty }(-s)^{l}\left[ \sideset{}{'}{\sum}_{k=1}^{[q/2]}\frac{%
(-1)^{k}s_{k}\sin \left( 2\pi k\alpha _{0}\right) }{\left[
l^{2}+(rs_{k}/w_{0})^{2}\right] ^{5/2}}\right.  \notag \\
&&\left. +\frac{q}{\pi }\int_{0}^{\infty }dx\frac{g(q,\alpha _{0},2x)}{\cosh
(2qx)-\cos (q\pi )}\frac{\cosh x}{\left[ l^{2}+(r\cosh x/w_{0})^{2}\right]
^{5/2}}\right] .  \label{J2Lm0}
\end{eqnarray}%
Massless fermionic field is conformally invariant and the expression (\ref%
{J2Lm0}) is conformally related to the corresponding result in the Minkowski
bulk: $\langle j_{\phi }\rangle _{\mathrm{b}}=(w/a)^{5}\langle j_{\phi
}\rangle _{\mathrm{b}}^{\mathrm{(M)}}$. Here, $\langle j_{\phi }\rangle _{%
\mathrm{b}}^{\mathrm{(M)}}$ is the boundary-induced current density in the
region between two boundaries located at $w=0$ and $w=w_{0}$ in
(4+1)-dimensional Minkowski spacetime with a cosmic string (the line element
is given by the expression in the brackets of (\ref{ds2})). The boundary $%
w=0 $ in the Minkowskian problem is the conformal image of the AdS boundary.
The vacuum polarization effects induced by planar boundaries orthogonal to
cosmic string in the Minkowski bulk have been investigated in \cite%
{Beze11,Beze12,Beze13b,Beze18b}.

\subsection{Asymptotic and numerical analysis}

We start the asymptotic analysis considering the points near the string,
assuming that $r\ll w_{0}-w$. By taking into account that in this region the
argument of the Bessel function in (\ref{J2L5}) is small, for the leading
order term we find%
\begin{equation}
\langle j_{\phi }\rangle _{\mathrm{b}}\approx \frac{erh_{-1}(q,\alpha _{0})}{%
4\pi ^{2}a^{5}}\int_{0}^{\infty }dx\,x^{5}\frac{K_{\nu _{1}}(xw_{0}/w)}{%
I_{\nu _{1}}(xw_{0}/w)}\left[ I_{\nu _{1}}^{2}(x)-I_{\nu _{2}}^{2}(x)\right]
,  \label{j2Lnear}
\end{equation}%
and the brane induced VEV linearly vanishes on the string. At large
distances from the string we use the representation (\ref{j2L2}). The
dominant contribution comes from the lowest eigenmode for $p_{i}$ and for $%
q\geq 2$ the VEV of the current density is suppressed by the exponential
factor $\exp [-2rp_{1}\sin (\pi /q)/w_{0}]$. For $q<2$ the suppression is
stronger, by the factor $e^{-2rp_{1}/w_{0}}$. Recall that in the R-region
the decay of the vacuum current at large distances from the string is
power-law for both massless and massive fields.\

For points close to the AdS boundary, $w/w_{0}\ll 1$, the dominant
contribution in the integral over $p$ given in (\ref{J2L5}) comes from the
region of integration where the argument of the functions $I_{\nu _{l}}(pw)$
is small. By using the corresponding asymptotic expression \cite{Abra}, to
the leading order one has
\begin{eqnarray}
\langle j_{\phi }\rangle _{\mathrm{b}} &\approx &\frac{%
se(w/w_{0})^{2ma+5}w_{0}^{2}}{2^{2ma}\pi ^{2}a^{5}\Gamma ^{2}(ma+1/2)r}%
\int_{0}^{\infty }dx\,x^{2ma+2}\frac{K_{\nu _{1}}(x)}{I_{\nu _{1}}(x)}
\notag \\
&&\times \left[ \sideset{}{'}{\sum}_{k=1}^{[q/2]}\frac{(-1)^{k}}{s_{k}}\sin
\left( 2\pi k\alpha _{0}\right) J_{2}(2xrs_{k}/w_{0})\right.  \notag \\
&&\left. +\frac{q}{\pi }\int_{0}^{\infty }dx\frac{g(q,\alpha _{0},2x)/\cosh x%
}{\cosh (2qx)-\cos (q\pi )}J_{2}(2xr\cosh x/w_{0})\right] .
\label{j2LnearAdSb}
\end{eqnarray}%
As we observe, the brane-induced current on the AdS boundary vanishes as $%
w^{2ma+5}$. Note that this suppression factor does not depend on the
parameter $s$. The expression in the right-hand side of (\ref{j2LnearAdSb})
also describes the behavior of the current in the limit when the location of
the brane is close to the AdS horizon for fixed $w$. If in addition $%
w_{0}\gg r$, that expression is further simplified as
\begin{equation}
\langle j_{\phi }\rangle _{\mathrm{b}}\approx \frac{seh_{-1}(q,\alpha
_{0})(w/w_{0})^{2ma+5}r}{2^{2ma+1}\pi ^{2}a^{5}\Gamma ^{2}(ma+1/2)}%
\int_{0}^{\infty }dx\,x^{2ma+4}\frac{K_{\nu _{1}}(x)}{I_{\nu _{1}}(x)}.
\label{j2LnearHor}
\end{equation}

For the evaluation of the current density on the brane it is more convenient
to use the representation (\ref{j2L2}) directly putting $w=w_{0}$:
\begin{eqnarray}
\langle j_{\phi }\rangle |_{w=w_{0}} &=&-\frac{ew_{0}^{2}}{\pi ^{2}a^{5}r}%
\sum_{i=1}^{\infty }p_{i}^{2}\left[ \sideset{}{'}{\sum}_{k=1}^{[q/2]}\frac{%
(-1)^{k}}{s_{k}}\sin \left( 2\pi k\alpha _{0}\right) K_{2}\left(
2rp_{i}s_{k}/w_{0}\right) \right.  \notag \\
&&\left. +\frac{q}{\pi }\int_{0}^{\infty }dx\frac{g(q,\alpha _{0},2x)/\cosh x%
}{\cosh (2qx)-\cos (q\pi )}K_{2}\left( 2rp_{i}\cosh x/w_{0}\right) \right] \
.  \label{j2LonBr}
\end{eqnarray}%
The consideration of the Minkowskian limit is similar to that of the
R-region. The corresponding result is given by (\ref{j2M}) with $y-y_{0}$
replaced by $y_{0}-y$. Of course, in the Minkowskian limit the VEV is
symmetric with respect to the brane. For the AdS bulk, though the local
geometrical characteristics are the same in the R- and L-regions, the
extrinsic curvature tensor, $K_{\mu \nu }$, for the brane under
consideration is different for those regions ($K_{\mu \nu }=-g_{\mu \nu }/a$
in the R-region and $K_{\mu \nu }=g_{\mu \nu }/a$ in the L-region). As a
consequence, the physical properties of the vacuum state, in particular, the
current densities, differ in those regions. The limiting values of the
current densities on the brane are different for the R- and L-regions. The
discontinuity is related to the idealized model for the brane with zero
thickness that constraints all the modes of the vacuum fluctuations. This
idealization leads, for example, to surface divergences in the fermion
condensate and in the VEV of the energy-momentum tensor. In more realistic
models, the fluctuations with high frequencies are relatively insensitive to
the presence of the brane. We expect that the formulas for the current
density given above will correctly approximate the corresponding results in
more realistic models for distances from the brane larger than the cutoff
wavelength. The latter is determined by the specific model for the brane.

\begin{figure}[tbph]
\begin{center}
\begin{tabular}{cc}
\epsfig{figure=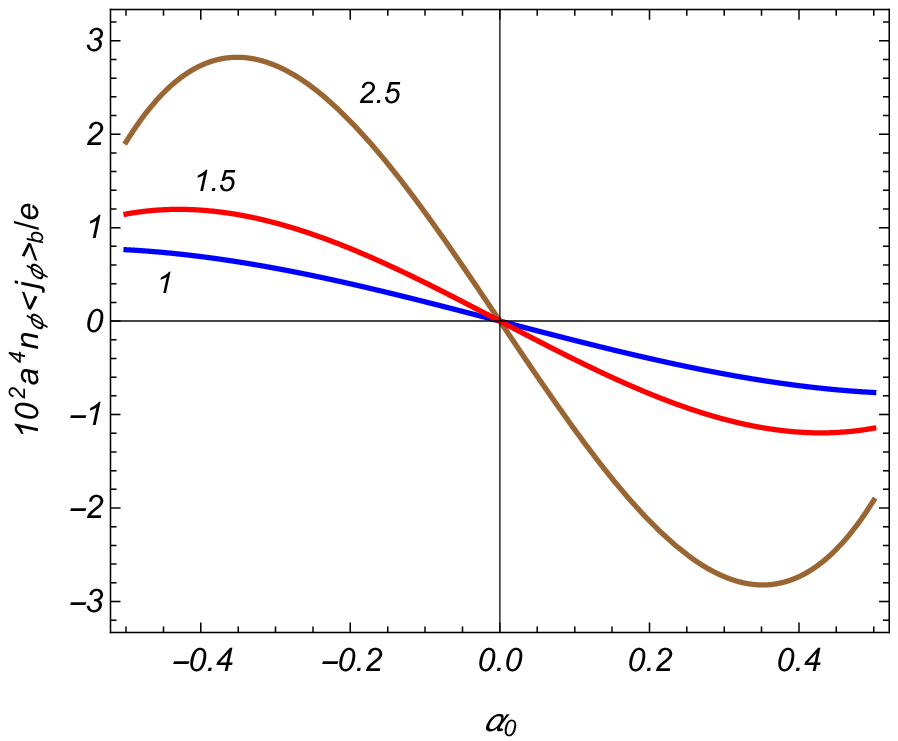,width=7.cm,height=5.5cm} & \quad %
\epsfig{figure=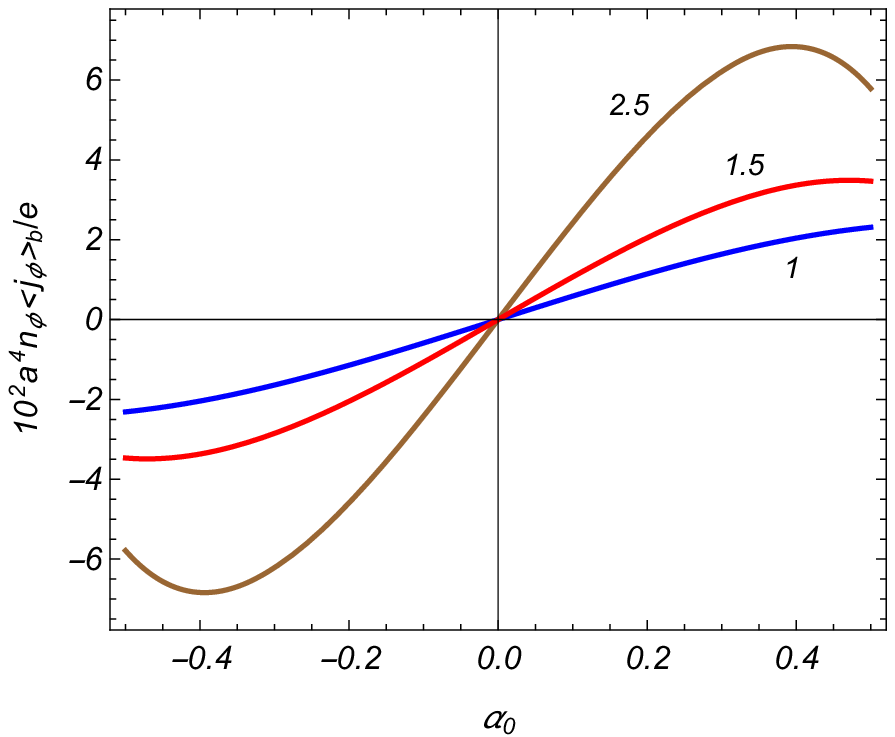,width=7.cm,height=5.5cm}%
\end{tabular}%
\end{center}
\caption{The brane-induced current density in the L-region versus the
parameter $\protect\alpha _{0}$. Graphs are plotted for $ma=1$, $w/w_{0}=0.8$%
, $r/w_{0}=0.5$. }
\label{fig7}
\end{figure}

Now we turn to the results of numerical investigations for the current
density in the L-region. As before, the numbers near the curves present the
values of the parameter $q$. The left and right panels in figures correspond
to fields with $s=1$ and $s=-1$, respectively. In figure \ref{fig7} the
brane-induced current density is plotted as a function of $\alpha _{0}$ for $%
ma=1$, $w/w_{0}=0.8$, $r/w_{0}=0.5$. For these values of the parameters the
absolute value of the current density increases with increasing $q$.
However, this is not a general feature. For points too close to the string
we have an opposite behavior (see figure \ref{fig9} below).

\begin{figure}[tbph]
\begin{center}
\begin{tabular}{cc}
\epsfig{figure=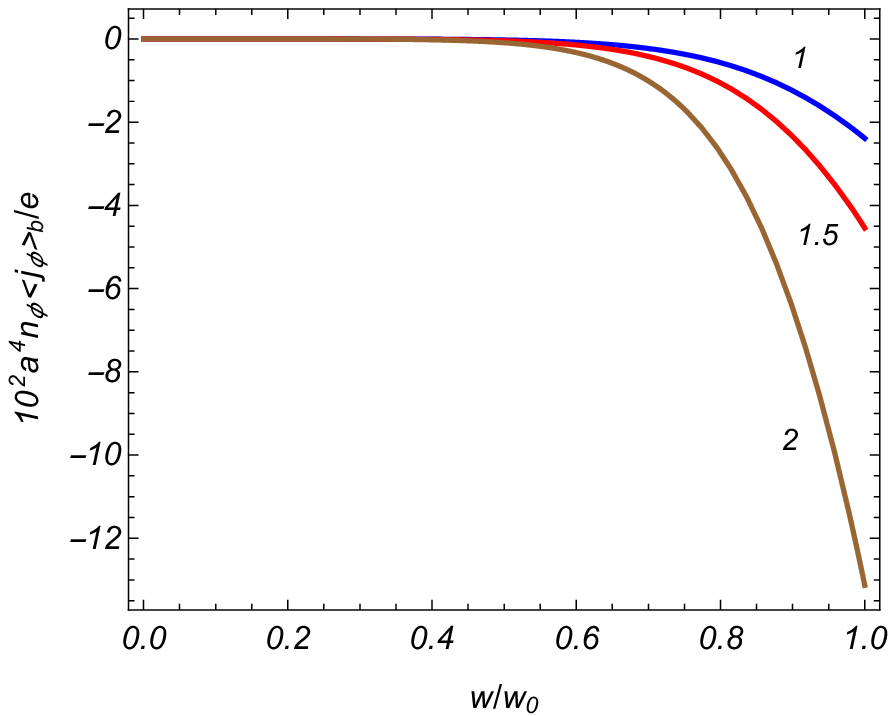,width=7.cm,height=5.5cm} & \quad %
\epsfig{figure=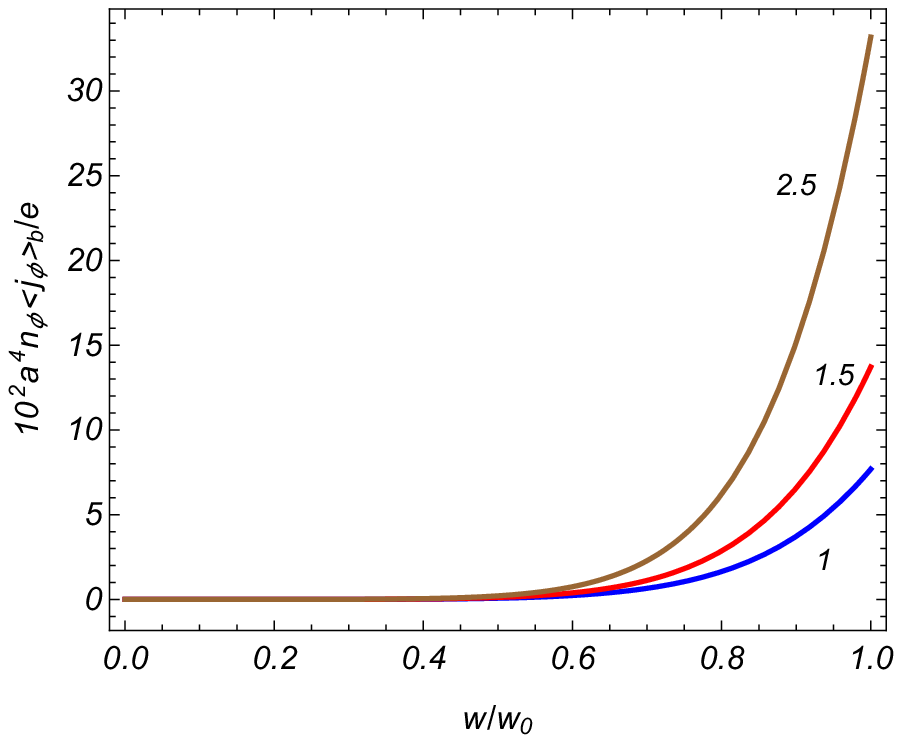,width=7.cm,height=5.5cm}%
\end{tabular}%
\end{center}
\caption{The brane contribution in the current density as a function of $%
w/w_{0}$ for $ma=1$, $\protect\alpha _{0}=0.3$, $r/w_{0}=0.5$. }
\label{fig8}
\end{figure}

Figure \ref{fig8} displays the dependence of the current density on the
ratio $w/w_{0}$ for fixed $ma=1$, $\alpha _{0}=0.3$, $r/w_{0}=0.5$. As it
has been already shown by the asymptotic analysis, the current density tends
to zero on the AdS boundary like $w^{2ma+5}$ for both cases $s=1$ and $s=-1$%
.

The brane-induced current density versus the distance from the string is
depicted in figure \ref{fig9} for $ma=1$, $\alpha _{0}=0.3$, $w/w_{0}=0.8$.
Recall that in the L-region the suppression of the current density at large
distances from the string is stronger compared to the behavior in the
R-region. One has an exponential decay insted of power-law fall-off for the
R-region.
\begin{figure}[tbph]
\begin{center}
\begin{tabular}{cc}
\epsfig{figure=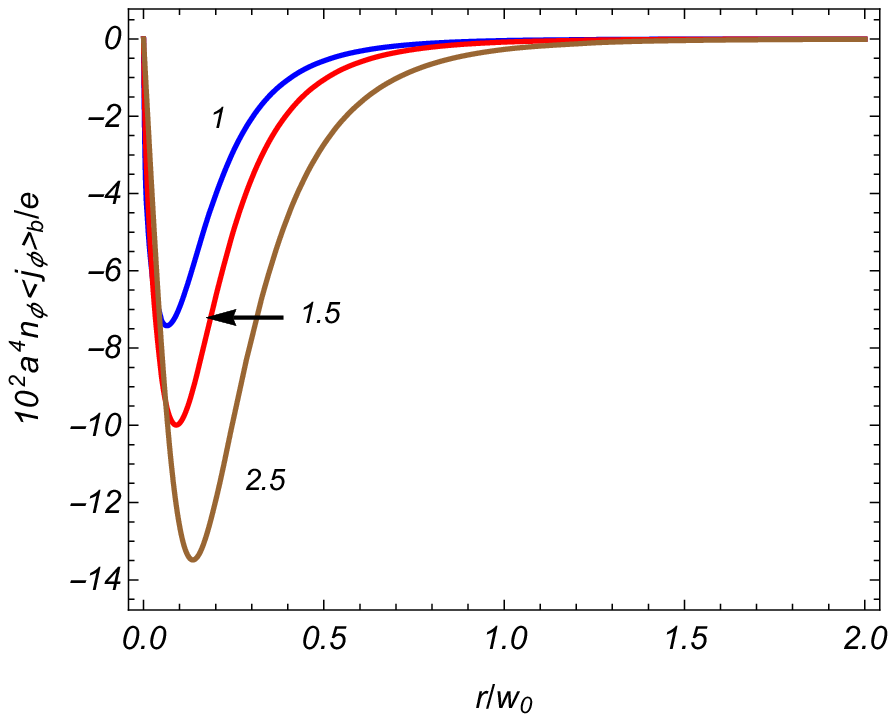,width=7.cm,height=5.5cm} & \quad %
\epsfig{figure=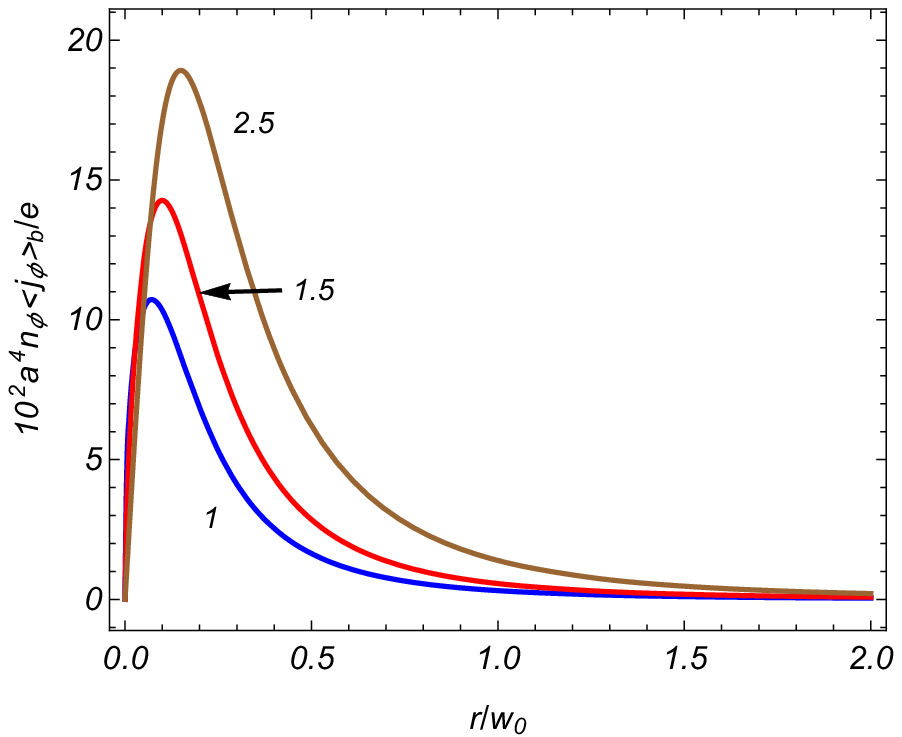,width=7.cm,height=5.5cm}%
\end{tabular}%
\end{center}
\caption{The dependence of the brane-induced current density on the distance
from the string for $ma=1$, $\protect\alpha _{0}=0.3$, $w/w_{0}=0.8$. }
\label{fig9}
\end{figure}

\begin{figure}[tbph]
\begin{center}
\begin{tabular}{cc}
\epsfig{figure=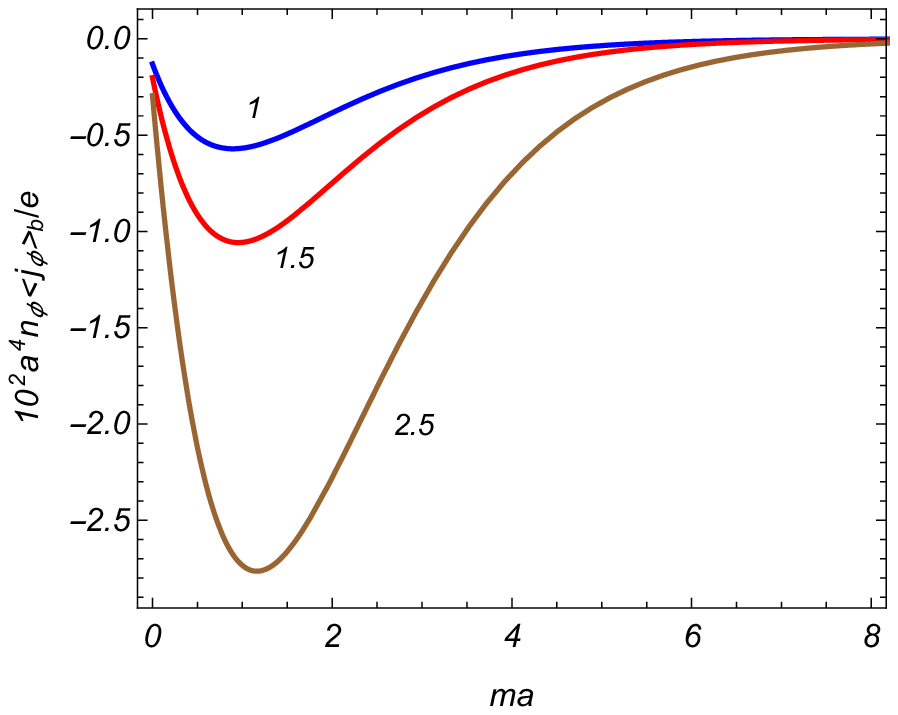,width=7.cm,height=5.5cm} & \quad %
\epsfig{figure=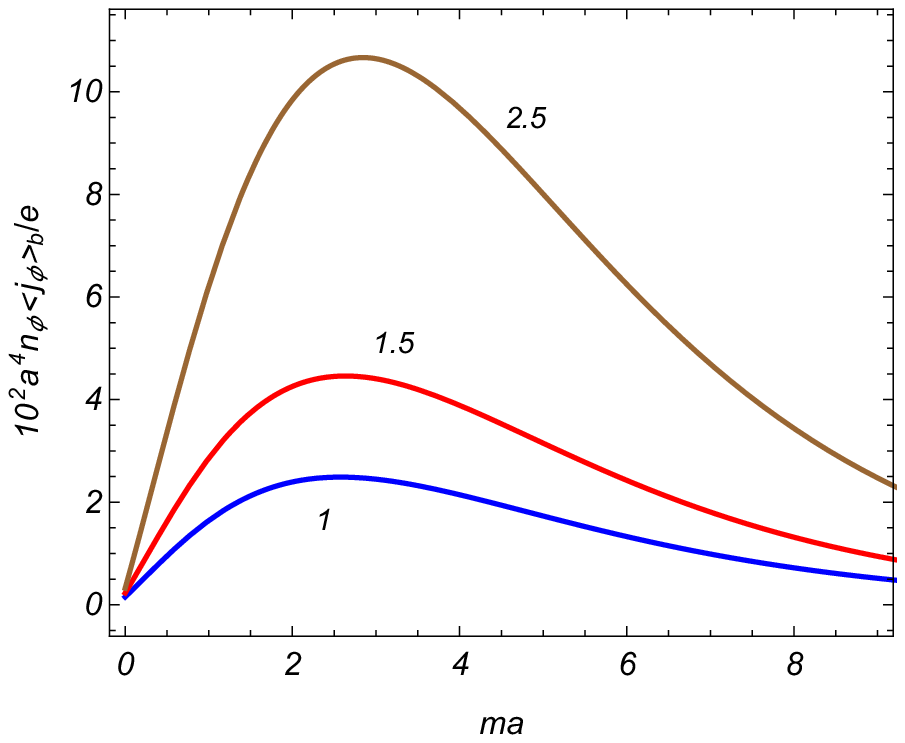,width=7.cm,height=5.5cm}%
\end{tabular}%
\end{center}
\caption{The brane-induced current density in the L-region versus the field
mass for $\protect\alpha _{0}=0.3$, $w/w_{0}=0.8$, $r/w_{0}=0.5$.}
\label{fig10}
\end{figure}

Figure \ref{fig10} presents the brane-induced current density as a function
of the mass for the values $\alpha _{0}=0.3$, $w/w_{0}=0.8$, $r/w_{0}=0.5$.
Unlike the R-region, the current density for massless fields does not
vanish. In the case $s=1$ (left panel), the corresponding values of the
combination $10^{2}a^{4}\langle j_{\phi }\rangle _{\mathrm{b}}/e$, plotted
in figure \ref{fig10}, are equal $-0.131,-0.201,-0.295$ for $q=1,1.5,2.5$,
respectively. For the field $s=-1$ (right panel) the respective values are
given by $0.153,0.23,0.323$. Again, we see that the current density is not a
monotonic function of the mass. At some intermediate value of the mass its
absolute value takes the maximum. For the example given in figure \ref{fig10}
the absolute value of the current density at the maximum is essentially
larger compared to the current density for a massless field.

\begin{figure}[tbph]
\begin{center}
\epsfig{figure=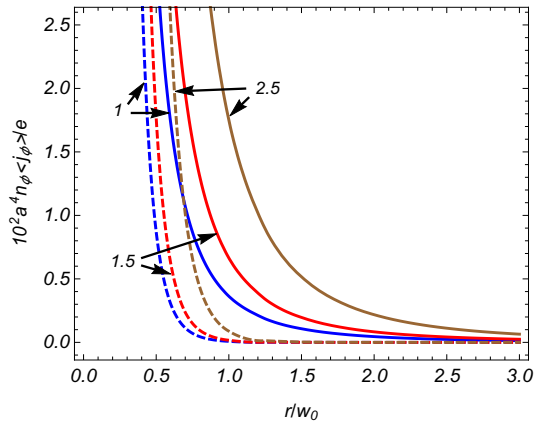,width=7.cm,height=5.5cm}
\end{center}
\caption{The same as in figure \protect\ref{fig5} for the total current
density in the L-region for $w/w_{0}=0.8$.}
\label{fig11}
\end{figure}

Finally, in figure \ref{fig11} we plot the total current density versus the
distance from the string for $ma=1$, $\alpha _{0}=0.3$, $w/w_{0}=0.8$. Full
and dashed curves correspond to the fields with $s=-1$ and $s=1$,
respectively. Similar to the R-region, at large distances from the string
the current density for the first case is dominated by the brane-induced
contribution.

\section{Vacuum current for the second type of boundary condition}

\label{sec:BC2}

In this section we consider the brane-induced contribution to the VEV\ of
the azimuthal current density for the boundary condition
\begin{equation}
(1-i\gamma ^{\mu }n_{\mu })\psi (x)=0\ ,\ \ w=w_{0}\ ,  \label{BC2}
\end{equation}%
The latter differs from (\ref{MITbc}) by the sign of the term containing the
Dirac matrices. As it will be seen below both the boundary conditions (\ref%
{MITbc}) and (\ref{BC2}) appear in Randall-Sundrum type models as a
consequence of the $Z_{2}$-symmetry with respect to the brane. The
evaluation procedure for the condition (\ref{BC2}) is similar to that we
have described in the previous sections and the final results will be
presented only.

\subsection{R-region}

We start with the R-region. The corresponding mode functions are given by (%
\ref{FermMod}) where now%
\begin{equation}
W_{\nu _{l}}(pw)=G_{\nu _{1},\nu _{l}}(pw_{0},pw),\;l=1,2.  \label{Wbc2}
\end{equation}%
The expression for the normalization coefficient is obtained from (\ref{CR})
by the replacement $\nu _{2}\rightarrow \nu _{1}$ in the orders of the
Bessel and Neumann functions. The mode-sum for the VEV of the azimuthal
current density is expressed as
\begin{eqnarray}
\langle j_{\phi }\rangle &=&-\frac{eqw^{6}}{4\pi ^{2}a^{5}}\int_{0}^{\infty
}d\lambda \int_{0}^{\infty }dp\int_{-\infty }^{+\infty }dk_{z}\sum_{j=\pm
1/2,...}\epsilon _{j}\frac{\lambda ^{2}p}{E}  \notag \\
&&\times J_{\beta _{j}}(\lambda r)J_{\beta _{j}+\epsilon _{j}}(\lambda r)%
\frac{G_{\nu _{1},\nu _{1}}^{2}(pw_{0},pw)+G_{\nu _{1},\nu
_{2}}^{2}(pw_{0},pw)}{J_{\nu _{1}}^{2}(pw_{0})+Y_{\nu _{1}}^{2}(pw_{0})},
\label{jbc2}
\end{eqnarray}%
with $w\geq w_{0}$. It is decomposed as (\ref{J2dec}) with the brane-induced
contribution
\begin{eqnarray}
\langle j_{\phi }\rangle _{\mathrm{b}} &=&\frac{ew^{6}}{2\pi ^{2}a^{5}r}%
\int_{0}^{\infty }dp\,p^{3}\frac{I_{\nu _{1}}(pw_{0})}{K_{\nu _{1}}(pw_{0})}%
\left[ K_{\nu _{1}}^{2}(pw)-K_{\nu _{2}}^{2}(pw)\right]  \notag \\
&&\times \left[ \sideset{}{'}{\sum}_{k=1}^{[q/2]}\frac{(-1)^{k}}{s_{k}}\sin
\left( 2\pi k\alpha _{0}\right) J_{2}(2prs_{k})\right.  \notag \\
&&\left. +\frac{q}{\pi }\int_{0}^{\infty }dx\,\frac{g(q,\alpha
_{0},2x)/\cosh x}{\cosh (2qx)-\cos (q\pi )}J_{2}(2pr\cosh x)\right] .
\label{jbc22}
\end{eqnarray}%
Now comparing these formulas with (\ref{j2R}) and (\ref{J2R5}), we see that
the VEVs of the current density for $s=\pm 1$ and for the boundary condition
(\ref{BC2}) coincide with the VEVs for the boundary condition (\ref{MITbc})
and for $s=\mp 1$.

\subsection{L-region}

In the L-region the mode functions for the boundary condition (\ref{BC2})
are given by (\ref{FermMod}) with $W_{\nu }(pw)=J_{\nu }(pw)$. The
eigenvalues $p_{i}$ of the quantum number $p$ are determined from the
boundary condition and are roots of the equation $J_{\nu _{2}}(pw_{0})=0$.
The normalization coefficient is given by (\ref{CL}) with the replacement $%
\nu _{2}\rightarrow \nu _{1}$ in the order of the Bessel function. For the
VEV\ of the azimuthal current density one gets%
\begin{eqnarray}
\langle j_{\phi }\rangle &=&-\frac{eqw^{6}}{2\pi ^{2}a^{5}w_{0}^{2}r}%
\int_{0}^{\infty }d\lambda \int_{-\infty }^{+\infty
}dk_{z}\sum_{j}\sum_{i=1}^{\infty }\epsilon _{j}\frac{\lambda ^{2}}{E}
\notag \\
&&\times \frac{J_{\beta _{j}}(\lambda r)J_{\beta _{j}+\epsilon _{j}}(\lambda
r)}{J_{\nu _{1}}^{2}\left( p_{i}\right) }\left[ J_{\nu
_{1}}^{2}(p_{i}w/w_{0})+J_{\nu _{2}}^{2}(p_{i}w/w_{0})\right] \ .
\label{jbc2L}
\end{eqnarray}%
In a way similar to that we have used in section \ref{sec:Lreg}, by using
the summation formula that is obtained from (\ref{AP}) by the replacements $%
\nu _{2}\rightarrow \nu _{1}$, $\nu _{1}\rightarrow \nu _{2}$, for the
brane-induced contribution in the L-region we find%
\begin{eqnarray}
\langle j_{\phi }\rangle _{\mathrm{b}} &=&\frac{ew^{6}}{2\pi ^{2}a^{5}r}%
\int_{0}^{\infty }dp\,p^{3}\frac{K_{\nu _{2}}(pw_{0})}{I_{\nu _{2}}(pw_{0})}%
\left[ I_{\nu _{2}}^{2}(pw)-I_{\nu _{1}}^{2}(pw)\right]  \notag \\
&&\times \left[ \sideset{}{'}{\sum}_{k=1}^{[q/2]}\frac{(-1)^{k}}{s_{k}}\sin
\left( 2\pi k\alpha _{0}\right) J_{2}(2prs_{k})\right.  \notag \\
&&\left. +\frac{q}{\pi }\int_{0}^{\infty }dx\,\frac{g(q,\alpha
_{0},2x)/\cosh x}{\cosh (2qx)-\cos (q\pi )}J_{2}(2pr\cosh x)\right] .
\label{jbc2L2}
\end{eqnarray}%
For a given $s$, the VEV of the azimuthal current density for the boundary
condition (\ref{BC2}) coincides with that for the condition (\ref{MITbc})
and for $s$ replaced by $-s$.

\section{Current density in $C$- and $P$-symmetric models}

\label{sec:CP}

In odd-dimensional spacetimes the Clifford algebra has two inequivalent
irreducible representations. For the spatial dimension $D=4$ we consider
here, the corresponding sets of the flat spacetime gamma matrices can be
taken in the form $\gamma _{(s)}^{(b)}=\{\gamma ^{(0)},\gamma ^{(1)},\gamma
^{(2)},\gamma ^{(3)},s\gamma ^{(4)}\}$, where $s=\pm 1$ specify the
representations and the matrices $\gamma ^{(b)}$ are obtained from the
corresponding matrices in (\ref{gam}) omitting the factors $w/a$. Note that
one has the relation $\gamma ^{(4)}=-\gamma ^{(0)}\gamma ^{(1)}\gamma
^{(2)}\gamma ^{(3)}$. We introduce fermionic fields $\psi _{(s)}$
corresponding to a given representation and having the Lagrangian $L_{(s)}=%
\bar{\psi}_{(s)}[i\gamma _{(s)}^{\mu }(\partial _{\mu }+\Gamma _{\mu
}^{(s)})-m]\psi _{(s)}$, where $\gamma _{(s)}^{\mu }=(w/a)\delta _{b}^{\mu
}\gamma _{(s)}^{(b)}$ are the curved spacetime gamma matrices and $\Gamma
_{\mu }^{(s)}$ is the respective spin connection. Note that, as it follows
from (\ref{gamGam}), the product $\gamma _{(s)}^{\mu }\Gamma _{\mu }^{(s)}$
is the same for $s=+1$ and $s=-1$. For massive fields, the separate
Lagrangians with a given $s$ are not invariant under the charge conjugation (%
$C$) and parity transformation ($P$). The models invariant under these
transformations in the absence of external gauge field can be constructed
considering the set of two fields with $s=+1$ and $s=-1$ and with the
Lagrangian $L=\sum_{s=\pm 1}L_{(s)}$. In these models the VEV of the total
current density, $\langle J^{\mu }\rangle $, is the sum of the VEVs $\langle
j_{(s)}^{\mu }\rangle $ corresponding to the separate fields: $\langle
J^{\mu }\rangle =\sum_{s=\pm 1}\langle j_{(s)}^{\mu }\rangle $, where $%
j_{(s)}^{\mu }=e\bar{\psi}_{(s)}\gamma _{(s)}^{\mu }\psi _{(s)}$. Note that
the matrices $\gamma _{(+1)}^{\mu }=\gamma ^{\mu }$ coincide with those we
have used in the previous sections.

Let us pass to a new set of fields $\psi _{(s)}^{\prime }$ by the
transformations $\psi _{(+1)}^{\prime }=\psi _{(+1)}$, $\psi _{(-1)}^{\prime
}=-\gamma ^{(4)}\psi _{(-1)}$. By taking into account that $\gamma
^{(4)}\gamma _{(-1)}^{\mu }\gamma ^{(4)}=\gamma ^{\mu }$ and $\gamma
_{(s)}^{\mu }\Gamma _{\mu }^{(s)}=\gamma ^{\mu }\Gamma _{\mu }$, the total
Lagrangian is presented in the form $L=\sum_{s=\pm 1}\bar{\psi}%
_{(s)}^{\prime }[i\gamma ^{\mu }(\partial _{\mu }+\Gamma _{\mu })-sm]\psi
_{(s)}^{\prime }$. For the fields $\psi _{(s)}^{\prime }$ the Dirac equation
is reduced to (\ref{Direq}). This shows that the parameter $s=\pm 1$ in the
previous discussion corresponds to different representations of the Clifford
algebra. For the current densities of the separate fields we get $%
j_{(s)}^{\mu }=e\bar{\psi}_{(s)}^{\prime }\gamma ^{\mu }\psi _{(s)}^{\prime
} $.

Now let us turn to the boundary conditions on the brane. First we consider
the case when both the fields $\psi _{(s)}$ obey the boundary condition (\ref%
{MITbc}):%
\begin{equation}
(1+i\gamma _{(s)}^{\mu }n_{\mu })\psi _{(s)}=0,\;w=w_{0}.  \label{MITbc2}
\end{equation}%
In terms of the new fields $\psi _{(s)}^{\prime }$ these conditions take the
form%
\begin{equation}
(1+si\gamma ^{\mu }n_{\mu })\psi _{(s)}^{\prime }=0,\;w=w_{0}.
\label{MITbc3}
\end{equation}%
We see that the field $\psi _{(+1)}^{\prime }$ obeys the field equation and
the boundary condition discussed in sections \ref{sec:Rreg} and \ref%
{sec:Lreg} for the case $s=1$. Hence, the brane-induced contributions in the
corresponding current densities in the R- and L-regions are given by (\ref%
{J2R5}) and (\ref{J2L5}) with $s=1$ and $\nu _{l}=ma+(-1)^{l}/2$. The field $%
\psi _{(-1)}^{\prime }$ obeys the field equation (\ref{Direq}) with $s=-1$
and the boundary condition (\ref{BC2}). As it has been discussed in the
previous section, the corresponding VEVs of the current density in the R-
and L-regions coincide with those considered in sections \ref{sec:Rreg} and %
\ref{sec:Lreg} for the case $s=1$ and for the condition (\ref{MITbc}).
Hence, we conclude that in the case of the boundary conditions (\ref{MITbc2}%
) the current densities $j_{(s)}^{\mu }$ coincide for the fields
corresponding to the representations $s=1$ and $s=-1$. The total current
density is obtained from the expressions given in sections \ref{sec:Rreg}
and \ref{sec:Lreg} with $s=1$ and with an additional factor 2.

From (\ref{MITbc3}) we see that the boundary conditions for the fields $\psi
_{(+1)}^{\prime }$ and $\psi _{(-1)}^{\prime }$ differ by the sign of the
terms with the Dirac matrices. We can consider the case where those terms
have the same sign and the transformed fields $\psi _{(s)}^{\prime }$ obey
the same boundary condition $(1+i\gamma ^{\mu }n_{\mu })\psi _{(s)}^{\prime
}=0$. That corresponds to the conditions $(1+si\gamma _{(s)}^{\mu }n_{\mu
})\psi _{(s)}=0$ for the initial fields on the brane $w=w_{0}$. With these
boundary conditions, the current density for the field with $s=1$ remains
the same as that for (\ref{MITbc2}) and is given by the expressions in
sections \ref{sec:Rreg} and \ref{sec:Lreg} with $s=1$. However, now the VEV $%
\langle j_{(-1)}^{\mu }\rangle $ is different. It is given by the
expressions from sections \ref{sec:Rreg} and \ref{sec:Lreg} with $s=-1$. In
this case the fields realizing two inequivalent irreducible representations
of the Clifford algebra give different contributions to the brane-induced
part in the VEV of the total current density.

\section{Applications to Randall-Sundrum model}

\label{sec:RS}

The results given above can be directly applied for the investigation of the
cosmic string induced effects in background of the Randall-Sundrum model
with a single brane (RSII model) \cite{Rand99,Rand99b} (for quantum effects
in higher-dimensional generalizations of RSII model see \cite{Saha20} and
references therein). In the presence of cosmic string perpendicular to the
brane, the corresponding background geometry contains two copies of the
R-region that are identified by the $Z_{2}$-symmetry with respect to the
brane located at $y=0$ (or at $w=w_{0}=a$ in terms of the coordinate $w$).
The corresponding line element is expressed as
\begin{equation}
ds^{2}=e^{-2|y|/a}\left( dt^{2}-dr^{2}-r^{2}d\phi ^{2}-dz^{2}\right)
-dy^{2}\ ,  \label{dsRS}
\end{equation}%
where, as before, $-\infty <y<+\infty $ and $0\leq \phi \leq \phi _{0}$.
Note that for an observer located on the brane $y=0$ the line element (\ref%
{dsRS}) is reduced to the standard line element for a cosmic string in
(3+1)-dimensional Minkowski spacetime. The boundary condition for the field $%
\psi (x)$ on the brane is obtained on the basis of the $Z_{2}$-symmetry (see
discussion in \cite{Bell18,Flac01b}).

Let $M$ be the matrix that relates the fields in the regions $-\infty <y<0$
and $0<y<+\infty $:
\begin{equation}
\psi (t,r,\phi ,-y,z)=M\psi (t,r,\phi ,y,z).  \label{psiZ2}
\end{equation}%
We require the invariance of the fermionic field action under the $Z_{2}$
identification of the points in those region. This requirement leads to the
following conditions on the matrix $M$:
\begin{equation}
\{\gamma ^{(0)},M\}=0,\;\{\gamma ^{(0)}\gamma ^{(3)},M\}=0,\;[\gamma
^{(0)}\gamma ^{(b)},M]=0,  \label{Mcond}
\end{equation}%
where $b=1,2,4$. By direct substitution it can be checked that these
relations are obeyed by the matrix $M=\varepsilon \gamma ^{(3)}$ with a
constant $\varepsilon $. For the latter from the relation $M^{2}=1$ we find $%
\varepsilon =\pm i$ and the transformation matrix has the form%
\begin{equation}
M=\pm \left( {%
\begin{array}{cc}
\sigma ^{3} & 0 \\
0 & -\sigma ^{3}%
\end{array}%
}\right) .  \label{MT}
\end{equation}%
Hence, we have two types of fermionic fields with the identification rule (%
\ref{psiZ2}) and with the upper and lower signs in (\ref{MT}).

On the brane at $y=0$ one gets the following boundary condition%
\begin{equation}
\left( 1-M\right) \psi (x)=0,\;w=a.  \label{BCrs}
\end{equation}%
Now we can see that the boundary condition (\ref{BCrs}) with the upper sign
in (\ref{MT}) coincides with the condition (\ref{MITbc}) and the boundary
condition (\ref{BCrs}) with the lower sign coincides with the condition (\ref%
{BC2}). Hence, for the fermionic field corresponding to the upper sign in (%
\ref{MT}), the expressions for the VEV of the current density induced by a
cosmic string in the RSII model is obtained from the formulas in section \ref%
{sec:Rreg} taking $w_{0}=a$ and adding an additional factor 1/2. For the
field with the lower sign in (\ref{MT}) the corresponding result is obtained
from the expression for the current density in section \ref{BC2} for the
R-region (again, with a factor 1/2). The appearance of the factor 1/2 is
related to that the integration in the normalization condition for the mode
functions goes over the interval $-\infty <y<+\infty $ instead of the
interval $0\leq y<+\infty $ for the R-region in the discussion above (with $%
y_{0}=0$).

The current density induced on the brane by vacuum fluctuations of a bulk
fermionic field is a source of magnetic fields surrounding the cosmic
string. These fields are among the distinctive features of
magnetic-flux-carrying strings. Note that the vacuum current density will
also be generated by quantum fluctuations of fields living on the brane (for
example, Standard Model fields in braneworld scenario). However, the
corresponding properties are different from those we have discussed above
(for fermionic currents in the geometry of a cosmic string on background of
(3+1)-dimensional Minkowski spacetime see \cite{Moha15,Beze10,Beze13d}). As
a consequence, the structure of magnetic fields around the cosmic string is
sensitive to the existence of extra dimensions.

\section{Conclusion}

\label{sec:Conc}

We have investigated the combined effects of the gravitational filed and a
brane on the vacuum current density for a charged fermionic field around a
magnetic-flux-carrying cosmic string. In order to have an exactly solvable
problem we have considered highly symmetric bulk and boundary geometries,
namely, AdS spacetime and a planar brane perpendicular to the cosmic string.
Two types of boundary conditions have been discussed on the brane. The first
one corresponds to the MIT bag boundary condition and the second one is
given by (\ref{BC2}). The current density for the field realizing the
representation $s=\pm 1$ and the boundary condition (\ref{BC2}) coincides
with the current density for the representation $s=\mp 1$ and for the
boundary condition (\ref{MITbc}) and we have described the main steps of the
evaluation procedure for the bag boundary condition only. The VEV of the
current density is expressed in the form of mode-sum over the complete set
of fermionic modes. The latter are given by (\ref{FermMod}). The brane
divides the space into two regions (R- and L-regions) with different
properties of the fermionic vacuum. In the R-region the function $W_{\nu
}(pw)$ in the mode functions is given by (\ref{WnuR}) and the spectrum of
the quantum number $p$ is continuous. In the L-region one has $W_{\nu
}(pw)=J_{\nu }(pw)$ and the eigenvalues for $p$ are determined by the zeros
of the function $J_{\nu _{1}}(pw_{0})$.

The only nonzero component of the vacuum current density corresponds to the
current along the $\phi $-direction (azimuthal current). It is decomposed
into brane-free and brane-induced parts. The brane-induced contributions in
the VEV of the azimuthal current density are given by the expressions (\ref%
{J2R5}) and (\ref{J2L5}) for the R- and L-regions, respectively. For a
massless field the brane-induced contribution vanishes in the R-region. The
current density is an odd periodic function of the magnetic flux along the
string axis, with the period equal the flux quantum. It is discontinuous at
half-integer values of the magnetic flux in units of flux \ quantum. The
corresponding limiting values are linear functions of the parameter $q$ and
are expressed as (\ref{J2Rhi}). In the special case $q=1$ the formulas
obtained provide the expressions for the current densities induced by
idealized magnetic flux tubes with zero thickness. In the limit of large
values for the curvature radius of the background spacetime, in the leading
order we have derived the expression (\ref{j2M}) for the fermionic current
density induced by a planar boundary, perpendicular to cosmic string in $%
(4+1)$-dimensional Minkowski spacetime.

To clarify the behavior of the azimuthal current density we have considered
various asymptotic regions of the parameters in the problem. Near the cosmic
string, for points not too close to the brane, the brane-induced
contribution is a linear function of the radial coordinate $r$ and vanishes
on the string. Near the string the brane-free part behaves as $\langle
j_{\phi }\rangle _{0}\propto 1/r^{4}$ and it dominates in the total VEV. At
large proper distances from the string, compared with the curvature radius
of the AdS spacetime, one has $r/w\gg 1$ and the brane-free VEV decays like $%
\langle j_{\phi }\rangle _{0}\propto \left( w/r\right) ^{2ma+4}$. The
behavior of the brane-induced contribution in the R-region at large
distances from the string is different for the fields realizing two
different irreducible representations of the Clifford algebra. For the
representation with $s=1$ the total current density at large distances is
dominated by the brane-free part. For the representation with $s=-1$ the
brane-induced contribution behaves as $\langle j_{\phi }\rangle _{\mathrm{b}%
}\propto (w/r)^{|2ma-1|+3}$ and it is dominant in the total VEV. In all
these cases, the decay of the current density on AdS bulk, as a function of
the proper distance from the string, is power-law for both massless and
massive fields. For massive fields this large-distance asymptotic in the
R-region is in clear contrast to that for the Minkowski bulk where the decay
is exponential. At large distances from the string the current density in
L-region is suppressed by the exponential factor $e^{-2rp_{1}\sin (\pi
/q)/w_{0}}$ for $q\geq 2$ and by the factor $e^{-2rp_{1}/w_{0}}$ for $q<2$,
with $p_{1}/w_{0}$ being the lowest eigenvalue for the quantum number $p$.

An important difference of the vacuum current density from the VEV of the
energy-momentum tensor is its finiteness on the brane. The corresponding
limiting values for the R- and L-region are directly obtained from the
representations (\ref{J2R3}), (\ref{j2L2}) and they are given by (\ref%
{J2RonBr}) and (\ref{j2LonBr}). At large distances from the brane, $%
y-y_{0}\gg a$, $w\gg r$, the brane-induced current density in the R-region
is suppressed by the factor $e^{-\left( 2ma+s\right) (y-y_{0})/a}$ for $%
ma>-s/2$. The suppression is stronger for the representation $s=1$. For $%
s=-1 $ and $ma<1/2$ the brane-induced current density tends to finite value
on the AdS horizon ($w\rightarrow \infty $), determined from (\ref{J2hor2}).
For a fixed location of the brane and in the limit $w\rightarrow 0$,
corresponding to points on the AdS boundary, the brane-induced VEV behaves
like $w^{2ma+5}$. The same is the case for the brane-free contribution.

In $(4+1)$-dimensional spacetime the Lagrangian for a massive fermionic
field realizing the irreducible representation of the Clifford algebra is
not invariant under charge conjugation and parity transformation. $C$- and $%
P $-invariant fermionic models are constructed by combining the fields $\psi
_{(s)}$, $s=\pm 1$, corresponding to two inequivalent sets of Dirac matrices
$\gamma _{(s)}^{\mu }$. By the transformation of the fields, the model with
two fields $\psi _{(s)}$ is reduced to the model with the fields $\psi
_{(s)}^{\prime }$ such that the new fields obey the equation (\ref{Direq}).
When the fields $\psi _{(s)}$ obey the same boundary conditions, the
boundary conditions for the fields $\psi _{(s)}^{\prime }$ differ by the
sign of the term containing the normal to the boundary. In this case the
VEVs of the current densities for separate fields $\psi _{(s)}$ coincide. In
the second case the boundary conditions are the same for the fields $\psi
_{(s)}^{\prime }$ and differ by the sign of the term containing the normal
for the fields $\psi _{(s)}$. In this case the vacuum currents differ and
the corresponding expressions are obtained from the formulas given in
sections \ref{sec:Rreg} and \ref{sec:Lreg} with $s=+1$ and $s=-1$.

In the Randall-Sundrum model with a single brane the boundary conditions on
a bulk fermionic field is dictated by the $Z_{2}$-symmetry. Depending on the
parity of the field, two boundary conditions are obtained at the location of
the brane. They correspond to the conditions (\ref{MITbc}) and (\ref{BC2})
in our discussion. Consequently, the VEV of the current density for a bulk
fermionic field, induced by magnetic-flux-carrying cosmic string RSII model,
is obtained from the results in sections \ref{sec:Rreg} and \ref{sec:Lreg}
with an additional factor 1/2, related to the presence of two copies of the
R-region.

For the convenience of the reader, in table \ref{table1} we summarize the
main formulas for the VEV of the azimuthal current density.
\begin{table}
\caption{Summary of the expressions for the vacuum current density.}
\label{table1}
\begin{center}
\begin{tabular}{|l|l|}
\hline
Brane-free part of the current density & (\ref{j2R0}),(\ref{j20}) \\ \hline
Boundary-free contribution in the Minkowski bulk & (\ref{j0M2}) \\ \hline
Boundary-induced contribution in the Minkowski bulk & (\ref{j2M}) \\ \hline
Total current density in the R-region & (\ref{J2R3}) \\ \hline
Brane-induced contribution to the current density in the R-region & (\ref%
{J2R5}) \\ \hline
Total current density in the L-region & (\ref{j2L2}) \\ \hline
Brane-induced contribution to the current density in the L-region & (\ref%
{J2L5}) \\ \hline
Brane-induced contribution to the current density in the L-region & (\ref%
{J2Lm0}) \\
for a massless field &  \\ \hline
\end{tabular}%
\end{center}
\end{table}

\section*{Acknowledgments}

This study was financed in part by the Coordena\c c\~ao de Aperfei\c
coamento de Pessoal de N\'ivel Superior - Brasil (CAPES) - Finance Code 001.
E.R.B.M is partially supported by CNPq under Grant no 301.783/2019-3.


\begin{thebibliography}{99}
\bibitem{Callan1990} C. G. Callan, Jr., and F. Wilczek, Nucl. Phys. \textbf{%
B340}, 366 (1990).

\bibitem{Fronsdal1974} C. Fronsdal, Phys. Rev. D \textbf{10}, 589 (1974).

\bibitem{Fronsdal1975} C. Fronsdal and R. B. Haugen, Phys. Rev. D \textbf{12}%
, 3810 (1975).

\bibitem{Avis1978} S. J. Avis, C. J. Isham and D. Storey, Phys. Rev. D
\textbf{18}, 3565 (1978).

\bibitem{Fradkin1984} E. S. Fradkin and A. A. Tseytlin, Nucl. Phys. \textbf{%
B234}, 472 (1984).

\bibitem{Sakai1984} N. Sakai and Y. Tanii, Phys. Lett. \textbf{B146}, 38
(1984).

\bibitem{Dullemond1985} C. Dullemond and E. van Beveren, J. Math. Phys.
\textbf{26}, 2050 (1985).

\bibitem{Burgess1985} C. P. Burgess and C. A. L\"{u}tken, Phys. Lett.
\textbf{B153}, 137 (1985)

\bibitem{Heidenreich1987} W. F. Heidenreich, Phys. Rev. D \textbf{36}, 1685
(1987).

\bibitem{Camporesi1991} R. Camporesi, Phys. Rev. D \textbf{43}, 3958 (1991).

\bibitem{Kamela1999} M. Kamela and C. P. Burgess, Can. J. Phys. \textbf{77},
85 (1999)

\bibitem{Goldberger} W. D. Goldberger and I. Z. Rothstein, Phys. Rev. Lett.
\textbf{89}, 131601 (2002).

\bibitem{Das2006} A. Das and G. V. Dunne, Phys. Rev. D \textbf{74}, 044029
(2006).

\bibitem{Ahar11} O. Aharony, D. Marolf and M. Rangamani, JHEP 02(2011)041.

\bibitem{Ahar13} O. Aharony, M. Berkooz, D. Tong and S. Yankielowicz, JHEP
02(2013)076.

\bibitem{Fuji14} I. Fujisawa and R. Nakayama, Nucl. Phys. B \textbf{886},
135 (2014).

\bibitem{Ambr15} V. E. Ambru\c{s} and E. Winstanley, Phys. Lett. B \textbf{%
749}, 597 (2015).

\bibitem{Kent15} C. Kent and E. Winstanley, Phys. Rev. D \textbf{91}, 044044
(2015).

\bibitem{Belo16} A. Belokogne, A. Folacci and J. Queva, Phys. Rev. D \textbf{%
94}, 105028 (2016).

\bibitem{Ambr17} V. E. Ambru\c{s} and E. Winstanley, Class. Quantum Grav.
\textbf{34}, 145010 (2017).

\bibitem{Ambr18} V. E. Ambru\c{s}, C. Kent and E. Winstanley, Int. J. Mod.
Phys. D \textbf{27}, 1843014 (2018).

\bibitem{Dapp18} C. Dappiaggi, H. Ferreira and A. Marta, Phys. Rev. D
\textbf{98}, 025005 (2018).

\bibitem{Morl20} T. Morley, P. Taylor and E. Winstanley, \textit{Quantum
field theory on global anti-de Sitter space-time with Robin boundary
conditions}, arXiv:2004.02704.

\bibitem{Kibble} T. W. Kibble, J. Phys. A \textbf{9}, 1387 (1976).

\bibitem{V-S} A. Vilenkin and E. P. S. Shellard, \textit{Cosmic Strings and
Other Topological Defects} (Cambridge University Press, Cambridge, England,
1994).

\bibitem{Hind95} M. B. Hindmarsh and T. W. B. Kibble, Rep. Prog. Phys.
\textbf{58}, 477 (1995).

\bibitem{Witt85} E. Witten, Phys. Lett. B \textbf{153}, 243 (1985).

\bibitem{Dvali1999} G. Dvali and S. H. Henry Tye, Phys. Lett. B\textbf{\ 450}%
, 72 (1999).

\bibitem{Tye2008} S. H. Henry Tye, Lect. Notes Phys. \textbf{737}, 949
(2008).

\bibitem{Cope10} E. J. Copeland and T. W. B. Kibble, Proc. R. Soc. A \textbf{%
466}, 623 (2010).

\bibitem{Cope11} E. J. Copeland, L. Pogosian and T.Vachaspati, Class.
Quantum Grav. \textbf{28}, 204009 (2011)

\bibitem{Cher15} D. F. Chernoff and S. H. Henry Tye, Int. J. Mod. Phys. D
\textbf{24}, 1530010 (2015).

\bibitem{Sarangi2002} S. Sarangi and S. H. Henry Tye, Phys. Lett. B\textbf{\
536}, 185 (2002).

\bibitem{Davis2001} S. C. Davis, Phys. Lett. B \textbf{499}, 179 (2001).

\bibitem{Davis2007} S. C. Davis, Phys. Lett. B \textbf{645}, 323 (2007).

\bibitem{Heydari-Fard} M. Heydari-Fard, H. Razmi and S. Y. Rokni, Class.
Quantum Grav. \textbf{30}, 165001 (2013).

\bibitem{Braga2005} N. R. F. Braga and C. N. Ferreira, J. High Energy Phys.
\textbf{03} (2005) 039.

\bibitem{Abdalla2007} M. C. B. Abdalla, M. E. X. Guimar\~{a}es and J. M. H.
da Silva, Phys. Rev. D \textbf{75}, 084028 (2007).

\bibitem{Abdalla2015} M. C. B. Abdalla, P. F. Carlesso and J. M. Hoff da
Silva, , Eur. Phys. J. Plus \textbf{75}, 432 (2015).

\bibitem{Ghe1} M. H. Dehghani, A. M. Ghezelbash and R. B. Mann, Nucl. Phys.
B \textbf{625}, 389 (2002).

\bibitem{Cristine} C. A. Ballon Bayona, C. N. Ferreira and V. J. Vasquez
Otoya, Class. Quantum Grav. \textbf{28}, 015011 (2011).

\bibitem{deMell:2011ji} E. R. Bezerra de Mello and A. A. Saharian, J. Phys.
A \textbf{45} 115402 (2012).

\bibitem{Beze13} E. R. Bezerra de Mello, E. R. Figueiredo Medeiros and A. A.
Saharian, Class. Quantum Grav. \textbf{30}, 175001 (2013).

\bibitem{Oliv19} W. Oliveira dos Santos, H. F. Mota and E. R. Bezerra de
Mello, Phys. Rev. D \textbf{99}, 045005 (2019).

\bibitem{Oliv20} W. Oliveira dos Santos, E. R. Bezerra de Mello and H. F.
Mota, Eur. Phys. J. Plus \textbf{135}, 27 (2020).

\bibitem{Oliveira} S. Bellucci, W. Oliveira dos Santos and E. R. Bezerra de
Mello, Eur. Phys. J. C \textbf{80}, 963 (2020).

\bibitem{Beze15a} E. R. Bezerra de Mello, A. A. Saharian and V. Vardanyan,
Phys. Lett. B \textbf{741}, 155 (2015).

\bibitem{Bell15} S. Bellucci, A. A. Saharian and V. Vardanyan, JHEP
11(2015)092.

\bibitem{Bell16} S. Bellucci, A. A. Saharian and V. Vardanyan, Phys. Rev. D
\textbf{93}, 084011 (2016).

\bibitem{Bell17} S. Bellucci, A. A. Saharian and V. Vardanyan, Phys. Rev. D
\textbf{96}, 065025 (2017).

\bibitem{Bell18} S. Bellucci, A. A. Saharian, D. H. Simonyan and V.
Vardanyan, Phys. Rev. D \textbf{98}, 085020 (2018).

\bibitem{Bell20} S. Bellucci, A. A. Saharian, H. G. Sargsyan and V.
Vardanyan, Phys. Rev. D \textbf{101}, 045020 (2020).

\bibitem{Bordag} M. Bordag and N. Khusnutdinov, Class. Quantum Grav. \textbf{%
13}, L41 (1996).

\bibitem{Mello2013} E. R. Bezerra de Mello and A. A. Saharian, Eur. Phys. J.
C \textbf{73} 2532 (2013).

\bibitem{Moha15} A. Mohammadi, E. R. Bezerra de Mello and A. A. Saharian, J.
Phys. A: Math. Theor. \textbf{48}, 185401 (2015).

\bibitem{Beze15} E. R. Bezerra de Mello, V. B. Bezerra, A. A. Saharian and
H. H. Harutyunyan, Phys. Rev. D \textbf{91}, 064034 (2015).

\bibitem{Mello} E. R. Bezerra de Mello, V. B. Bezerra, A.A. Saharian and
V.M. Bardeghyan, Phys. Rev. D \textbf{82}, 085033 (2010).

\bibitem{Prud2} A. P. Prudnikov, Yu. A. Brychkov and O. I. Marichev, \textit{%
Integrals and Series} (Gordon and Breach, New York, 1986), Vol. 2.

\bibitem{Abra} \textit{Handbook of Mathematical Functions}, edited by M.
Abramowitz and I. A. Stegun (Dover, New York, 1972).

\bibitem{Bell16c} S. Bellucci, E. R. Bezerra de Mello, E. Braganca and A. A.
Saharian, Eur. Phys. J. C \textbf{76}, 350 (2016).

\bibitem{Saha19} A. A. Saharian, E. R. Bezerra de Mello and A. A. Saharyan,
Phys. Rev. D \textbf{100}, 105014 (2019).

\bibitem{Bell20b} S. Bellucci, I. Brevik, A. A. Saharian and H. G. Sargsyan,
Eur. Phys. J. C \textbf{80}, 281 (2020).

\bibitem{Bell16b} S. Bellucci, A. A. Saharian and A. Kh. Grigoryan, Phys.
Rev. D \textbf{94}, 105007 (2016).

\bibitem{Aram:2007} A. A. Saharian, \textit{The generalized Abel-Plana
formula with applications to Bessel functions and Casimir effect},
arXiv:0708.1187 [hep-th].

\bibitem{Beze11} E. R. Bezerra de Mello and A. A. Saharian, Class. Quantum
Grav. \textbf{28}, 145008 (2011).

\bibitem{Beze12} E. R. Bezerra de Mello, A. A. Saharian and A. Kh.
Grigoryan, J. Phys. A: Math. Theor. \textbf{45}, 374011 (2012).

\bibitem{Beze13b} E. R. Bezerra de Mello, A. A. Saharian and S. V. Abajyan,
Class. Quantum Grav. \textbf{30}, 015002 (2013).

\bibitem{Beze18b} E. R. Bezerra de Mello, A. A. Saharian and S. V. Abajyan,
Phys. Rev. D \textbf{97}, 085023 (2018).

\bibitem{Rand99} L. Randall and R. Sundrum, Phys. Rev. Lett. \textbf{83},
3370 (1999).

\bibitem{Rand99b} L. Randall and R. Sundrum, Phys. Rev. Lett. \textbf{83},
4690 (1999).

\bibitem{Saha20} A. A. Saharian, Universe \textbf{6}, 181 (2020).

\bibitem{Flac01b} A. Flachi, I.G. Moss and D.J. Toms, Phys. Rev. D \textbf{64%
}, 105029 (2001).

\bibitem{Beze10} E. R. Bezerra de Mello, Class. Quantum Grav. \textbf{27},
095017 (2010).

\bibitem{Beze13d} E. R. Bezerra de Mello and A. A. Saharian, Eur. Phys. J. C
\textbf{73}, 2532 (2013).
\end{thebibliography}
\end{document}